\documentclass[aps,prd,onecolumn,superscriptaddress,floats,amsfonts,amsmath,amssymb,linenumbers,12pt,nofootinbib]{revtex4}
\usepackage{graphicx}
\usepackage{epstopdf}
\usepackage{hyperref}
\usepackage[active]{srcltx}
\usepackage{bm}
\usepackage[]{latexsym}

\usepackage{color}

\usepackage{caption}
\usepackage{soul}

\usepackage{float}

\usepackage{lineno}
\usepackage{comment}
\usepackage[percent]{overpic}
\usepackage{feynmp-auto}
\newcommand{\marrow}[5]{%
	\fmfcmd{style_def marrow#1
		expr p = drawarrow subpath (1/4, 3/4) of p shifted 6 #2 withpen pencircle scaled 0.4;
		label.#3(btex #4 etex, point 0.5 of p shifted 6 #2);
		enddef;}
	\fmf{marrow#1,tension=0}{#5}}
\newcommand{\Marrow}[7]{%
	\fmfcmd{style_def marrow#1
		expr p = drawarrow subpath (1/4, 3/4) of p shifted #6 #2 withpen pencircle scaled 0.4;
		label.#3(btex #4 etex, point 0.5 of p shifted #7 #2);
		enddef;}
	\fmf{marrow#1,tension=0}{#5}}
\usepackage{slashed}

\usepackage{xcolor}
\definecolor{mypurple}{RGB}{255, 0, 255}
\definecolor{mygreen}{RGB}{0, 200, 0}
\definecolor{mylightblue}{RGB}{0, 200, 200}
\begin{document}
%\linenumbers

\restylefloat{figure}

%\title{Constraining the Gauged $U(1)_{L_\mu-L_\tau}$ Model by Supernova Neutrino Observation}
\title{SN1987A constraints to BSM models with extra neutral bosons near the trapping regime: 
$U(1)_{L_\mu-L_\tau}$ model as an illustrative example}

\author{Kwang-Chang Lai}
\affiliation{Center for General Education, Chang Gung University, Kwei-Shan, Taoyuan, 333, Taiwan}

\author{Chun Sing Jason Leung}
\affiliation{Institute of Physics, National Yang Ming Chiao Tung University, Hsinchu, 300, Taiwan}
\affiliation{Department of Physics, National Tsing Hua University, Hsinchu, 300, Taiwan}

\author{Guey-Lin Lin}
\affiliation{Institute of Physics, National Yang Ming Chiao Tung University, Hsinchu, 300, Taiwan}

\begin{abstract}

New physics beyond the Standard Model (BSM) with an extra neutral boson can be constrained from the observation of SN1987A, since productions of this neutral boson in a supernova (SN) could accelerate the SN cooling and potentially lead to a period of the neutrino burst incompatible with the observation.% for certain ranges of model parameters. 
The constraint to the model is formulated by the condition $L_{\rm NB}\leq 3\times 10^{52}$ erg/s according to G. Raffelt with $L_{\rm NB}$ the luminosity of BSM neutral boson. Computing the above luminosity in the large coupling case, the so-called trapping regime, is non-trivial since the luminosity is a competition between the large production rate and the efficient absorption or decay rate of the neutral boson. We illustrate such a subtlety using $U(1)_{L_\mu-L_\tau}$ model as an example where the $Z^{\prime}$ luminosity, $L_{Z^{\prime}}$, from the neutrinosphere is calculated.               
 
We calculate $Z'$ production, absorption, and decay rates through pair-coalescence, semi-Compton, loop-bremsstrahlung from proton-neutron scattering, and their inverse processes in a benchmark SN simulation with muons. We point out that, as the coupling constant $g_{Z'}$ increases, $L_{Z^{\prime}}$ shall be approaching to a constant plateau value for a given $m_{Z'}$ instead of monotonically decreasing down to zero as obtained in the previous literature. 
%As a result, the trapping limits for the parameters of $U(1)_{L_\mu-L_\tau}$ model are modified. 
We demonstrate that this plateau phenomenon can be understood by physical arguments and justified by numerical calculations. With a different result on $L_{Z^{\prime}}$ from the previous one, we discuss impacts on the constraints to $U(1)_{L_\mu-L_\tau}$ parameter space by SN1987A. The implication of our result to the similar constraint on a generic BSM model with an extra neutral boson is also discussed.      
%We stress that the plateau behavior of $Z'$ luminosity in the large coupling limit should also occur for other new physics models that introduce an additional light neutral boson. 

\end{abstract}

\maketitle

\section{Introduction}

Detecting neutrinos from SN1987A in the Large Magellanic Cloud confirms the paradigm of core-collapse supernovae and provides useful information on particle interactions and neutrino properties. The inferred energy release through neutrinos is about $3\times 10^{53}$ erg, agreeing well with the difference between the gravitational binding energy of the progenitor and the remnant. This observation is useful for constraining new physics beyond the Standard Model (BSM), which contains an extra neutral boson. This stems from the fact that the extra neutral boson produced in the proto-neutron star (PNS) could carry away energies, which lead to an accelerated cooling of PNS and consequently a shorter period of neutrino emissions.      
To set quantitative constraints on BSM models based upon the possible accelerated cooling of PNS, a simple criterion provided by Raffelt~\cite{Raffelt:1996wa} on the luminosity of the extra neutral boson can be applied. 
Note that the Raffelt criterion is not the only approach that 
can extract model constraints from supernovae observations.
%utilized SN. 
In addition to the Raffelt criterion, sensitivity analyses based on the absence of neutrino-nucleus or neutrino-lepton scattering events in experimental setups, such as Refs.~\cite{Akita:2023iwq, Fiorillo:2022cdq, Jegerlehner:1996kx, Mirizzi:2005tg}, can also provide model constraints.
%provide even more stringent constrains
%as experimental sensitivity improves over time. 
Since our current objective is to elucidate calculations near the trapping regime, we shall %initially 
adhere to the Raffelt criterion in this work and %may 
consider incorporating a more comprehensive experimental detector setup in future works.

The application of the Raffelt criterion can be divided into two separated cases: the free streaming limit and the trapping limit. The free streaming constraint requires that the energy carried away by BSM particle per unit time should not be comparable to or exceed the neutrino luminosity, which is $3\times 10^{52}$ erg s$^{-1}$ at post-bounce time $t_{\rm pb}=1$ s. Otherwise, it would significantly reduce the duration of the neutrino burst and contradict what has been observed for SN1987A. In this case the coupling constant between the BSM boson and SM particles should be smaller than a critical value. On the other hand, if the above coupling constant is sufficiently large, the BSM boson can be reabsorbed by the surrounding matter or it can decay within PNS such that its energy can be reprocessed in the star. In terms of the optical depth $\tau$ of the BSM boson, this new boson would not be able to carry out energies provided $\tau\geq 1$, with the equal sign understood as the trapping limit. In other words, the condition $\tau\geq 1$ implies that the mean free path of the BSM boson is smaller than the distance between the production point of the BSM boson and the surface of the neutrinosphere such that the SN cooling time is unaffected.

Examples of using the Raffelt criterion with these two limits are given in earlier works \cite{Dent:2012mx, Zhang:2014wra, Kazanas:2014mca,  Rrapaj:2015wgs, Hardy:2016kme, Mahoney:2017jqk}. Nevertheless, taking 
$\tau=1$
as the trapping limit is just an approximation, and it underestimates the constraint. An unified expression for the luminosity, encompassing the production and reabsorption of BSM neutral bosons, can be derived for simultaneously determining both the free streaming and trapping constraints, as proposed by Chang {\it et al.}~\cite{Chang:2016ntp}. In this expression, the luminosity $L_{\rm NB}$ is determined by four parameters $(R_\nu, R_{\rm far}, m_{\rm NB}, g_{\rm NB})$, which are the radius of the neutrinosphere, the cutoff radius for reabsorption and decay processes, the BSM particle mass, and the coupling strength of the BSM boson to SM particles, respectively. 
The authors emphasize the importance of ensuring that the value of $R_{\rm far}$ is larger 
%smaller 
than $R_{\nu}$. For example it has been argued in ~\cite{Chang:2016ntp,Croon:2020lrf} that $R_{\rm far}$
can be taken as the neutrino gain radius $R_{\rm gain}\approx$ 100 km which is defined in the way that there are net productions of neutrinos between $R_{\nu}$ and $R_{\rm gain}$ while neutrino number decreases beyond $R_{\rm gain}$~\cite{Bethe:1992fq,Janka:2000bt}. In the case of $U(1)_{L_\mu-L_\tau}$, as $Z'$ boson is reabsorbed in the region between $R_{\nu}$ and $R_{\rm gain}$, its energy will be inherited by nucleons absorbing $Z^{\prime}$ through $n+p+Z^{\prime}\to n+p$. Such energies would then be returned to neutrinos in the neutrino production processes $e^-+p\to n + \nu_e$ and $e^++n\to p+\bar{\nu}_e$. 

While $Z^{\prime}$ attenuation between $R_{\nu}$ and $R_{\rm gain}$ does reduce $L_{Z^{\prime}}$ as seen in~\cite{Croon:2020lrf}, we stress that it is also necessary to consider the production of $Z^{\prime}$ in the same region for the consistency.  
In other words, for a consistent calculation, the production and attenuation of $Z^{\prime}$ should be considered within the same region, no matter it is the space within the neutrinosphere or a larger region up to the neutrino gain radius for instance. In this article, we shall demonstrate such type of calculations using $U(1)_{L_\mu-L_\tau}$ model as an example and consider the region within the neutrinosphere for obtaining $L_{Z^{\prime}}$. Extending the region to the neutrino gain radius does not change $L_{Z^{\prime}}$ much as the number densities of all relevant particles drop significantly outside the neutrinosphere. Furthermore, the results in~\cite{Croon:2020lrf} do include the $R_{\rm far}\to R_{\nu}$ case for $m_{Z^{\prime}}=0.1$ eV, which is a useful point for comparing two calculations. We shall see later that $L_{Z^{\prime}}$ in the trapping regime approaches to a constant plateau value instead of decreasing to zero monotonically as $g_{Z^{\prime}}$ increases. A simple argument for this phenomenon will be given in later sections. We note that the result in~\cite{Croon:2020lrf}
with $m_{Z^{\prime}}=0.1$ eV does not exhibit this behaviour. 
%Such a phenomenon is due to the interplay between the production and attenuation of $Z^{\prime}$ in the large $g_{Z^{\prime}}$ limit.  
%instead of decreasing to zero monotonically.               

The $U(1)_{L_\mu-L_\tau}$ model is considered a minimal extension to the Standard Model of particle physics. In this model, it is possible that the new $U(1)$ local symmetry is spontaneously broken, which gives rise to a massive new gauge boson referred to as the $Z'$ boson~\cite{He:1990pn,He:1991qd,Ma:2001md,Baek:2001kca,Salvioni:2009jp,Heeck:2011wj,Harigaya:2013twa,Bauer:2018onh,Amaral:2021rzw,Greljo:2021npi,Langacker:2008yv,Foldenauer:2018zrz,Asai:2020qlp,Holst:2021lzm,Drees:2021rsg,Hapitas:2021ilr,Heeck:2022znj}. This neutral $Z'$ boson couples directly to the second generation leptons and neutrinos. At the one loop level, $Z'$ can mix with photon and $Z$ boson so that it couples indirectly to other fermions as well. Through these direct and indirect couplings, $Z'$ can be produced as well as reabsorbed in the PNS. As stated in ~\cite{Croon:2020lrf}, the mechanisms for producing $Z^{\prime}$ are pair coalescence processes $\nu \bar{\nu}\to Z^{\prime}$ and $\mu^+\mu^- \to Z^{\prime}$, the semi-Compton process $\mu^-\gamma\to \mu^- Z^{\prime}$, and the loop bremsstrahlung process $n+p\to n+p+Z^{\prime}$ where $Z^{\prime}$ couples to proton through its mixing with $\gamma$. On the other hand, $Z^{\prime}$ can also decay or be reabsorbed by inverse processes of the above.

The paper is organized as follows. In Section \ref{sec.02}, we present the Lagrangian of the model, describe the Raffelt criterion, provide an overall luminosity formula, and discuss the impact of $Z'$ boson emission on the energy release of SN1987A. Next, we explain the framework used for calculating the $Z'$ boson luminosity in Section \ref{sec.03}. In Sections \ref{sec.emissivity} and \ref{sec.attenuationlength}, we calculate the emissivity and attenuation factors for each individual reaction process.  Next, the overall luminosity of the $Z'$ boson in the PNS is calculated by integrating the emissivity and attenuation factors as discussed in Section \ref{sec.04}. The competition and cancellation between the emissivity and attenuation factors will be investigated in details. Subsequently, we present the model parameter space that can be excluded by SN1987A based upon our new calculation of $L_{Z^{\prime}}$, which approaches to a constant plateau value as $g_{Z^{\prime}}$ increases. This phenomenon can be seen through both numerical calculations and theoretical arguments. Finally, we conclude in Section \ref{sec.05}. The appendices provide comprehensive derivations of relevant formulas and provide detailed explanations for the plateau phenomenon.

\section{Impact of \texorpdfstring{$Z^{\prime}$}{} boson on SN1987A} \label{sec.02}

We first discuss the $U(1)_{L_\mu-L_\tau}$ extension of SM, which introduces the new boson $Z'$. The Lagrangian of the model is given by~\cite{He:1990pn,He:1991qd,Ma:2001md,Baek:2001kca,Salvioni:2009jp,Heeck:2011wj,Harigaya:2013twa,Bauer:2018onh,Amaral:2021rzw,Greljo:2021npi,Langacker:2008yv,Foldenauer:2018zrz,Asai:2020qlp,Holst:2021lzm,Drees:2021rsg,Hapitas:2021ilr,Heeck:2022znj}
\begin{eqnarray}
\mathcal{L}_{Z'} & = & \mathcal{L}_{\rm SM}-\frac{1}{4}Z'_{\mu\nu}Z'^{\mu\nu}-\frac{\epsilon}{2}Z'_{\mu\nu}Z^{\mu\nu}+             \frac{1}{2}m^2_{Z'}Z'_\mu Z'^\mu  \nonumber  \\
                 & + & g_{Z'}Z'_\mu\left(\bar{l}_1\gamma^\mu l_1-\bar{l}_2\gamma l_2+\bar{\mu}_R\gamma^\mu\mu_R+\bar{\tau}_R\gamma^\mu\tau_R\right), \label{Lmu-LtauLagrangian}
\end{eqnarray}
where $g_{Z'}$ is $U(1)_{L_\mu-L_\tau}$ gauge coupling, $l_1$ and $l_2$ denote the electroweak doublets for the left-handed leptons $(\mu_L,~\nu_{\mu,L})$ and $(\tau_L,~\nu_{\tau,L})$, while $\mu_R$ and $\tau_R$ are the electroweak singlets for the right-handed leptons. 
The mass of $Z'$ boson can be generated either through the Stueckelberg mechanism~\cite{Stueckelberg_1938,Ruegg:2003ps} or the spontaneous $U(1)_{L_\mu-L_\tau}$ symmetry breaking. For the purpose of comparing with Croon {\it et al}.~\cite{Croon:2020lrf}, we 
adopt Stueckelberg mechanism for generating the $Z'$ mass. The scenario of generating new boson mass by the so called dark Higgs has been discussed in~\cite{An:2013yua, Kahlhoefer:2015bea, Bell:2016fqf, Bell:2016uhg, Duerr:2016tmh, Bell:2017irk, Cui:2017juz}.  
In this scenario, there exist new effects to the $Z'$ luminosity. For instance, the $Z'$ luminosity could be enhanced through the decay $H' \rightarrow Z' + Z'$ when the dark Higgs is produced copiously. Additionally, dark Higgs production could serve as an extra SN cooling channel.

In this model, $Z'$ can only interact with muons, taus, and their corresponding neutrinos. Therefore, to produce a considerable amount of $Z'$ bosons, the core of PNS must contain a significant population of these leptons. In the core of PNS, the number of electrons overwhelms that of the positron to balance the positive charges of the protons. The chemical potential of these highly degenerate electrons is greater than the muon mass, $\mu_e>m_\mu$, and hence can be transferred into muons through SM processes~\cite{Bollig:2017lki}. In addition, a significant number of neutrinos and photons are generated through thermal pair production in the core, which in turn produces a considerable amount of muons. While muons are produced copiously, the production of tau leptons is suppressed due to the large tau lepton mass.

As supernova muons can be produced in large quantities within PNS, a considerable amount of $Z'$ bosons can be generated through $Z^{\prime}$ couplings to muons. If these $Z'$ bosons can free-stream out and carry a significant amount of energy away from the neutrinosphere, the energy that drives the production of supernova neutrinos will decrease accordingly. Therefore, the production of $Z'$ bosons will affect the neutrino emission of SN explosions and the overall PNS cooling duration. As the observations of SN1987A neutrinos are consistent with SN simulations having the neutrino emission as the only cooling mechanism, the $Z'$ boson emissions cannot dominate the cooling, which then leads to a constraint on the $(m_{Z^{\prime}}, g_{Z'})$ parameter space. Such a  constraint can be formulated as an upper limit for the $Z^{\prime}$ luminosity, referred to as the Raffelt criterion~\cite{Raffelt:1996wa} for the generic case. Raffelt's criterion based upon SN1987A observation states that
\begin{equation}\label{eq.RaffeltC}
L_{z'} < L_\nu = 3\times10^{52} ~~ {\rm erg/s}.
\end{equation}
To be consistent with observations, the energy carried away by $Z'$ must be considerably lower than that carried away by neutrinos.

We now describe the way we calculate the $Z'$ boson luminosity, $L_{Z'}$. 
%following the framework in \cite{Chang, Croon:2020lrf}. 
In the free-streaming region, where no re-absorption occurs and $Z'$ bosons can flee without attenuation, the $Z'$ boson luminosity can be directly inferred from its production rate $\Gamma_{\rm prod}$:
\begin{equation}\label{eq.ULum}
L_{Z'} = \int dV\int\frac{d^3k}{(2\pi)^3}\omega\Gamma_{\rm prod},
\end{equation}
where $\omega$ is the $Z'$ energy, and $V$ denotes the volume of the neutrinosphere. 

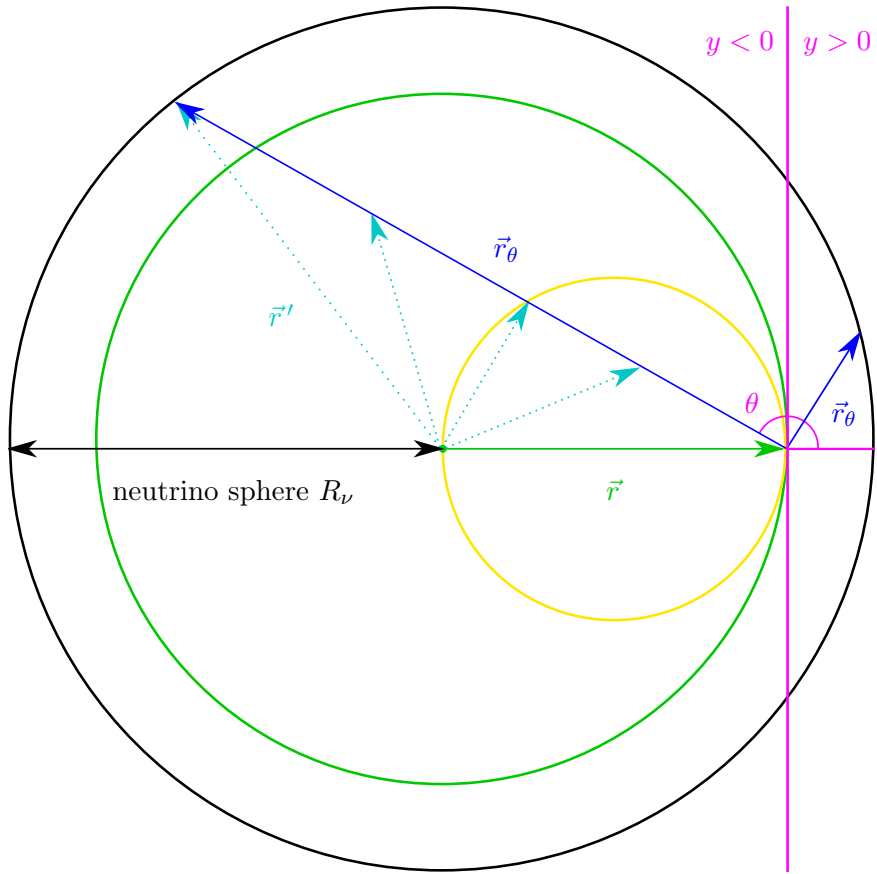
\begin{figure}[htbp]
\centering
	\begin{overpic}[width=0.7\columnwidth]{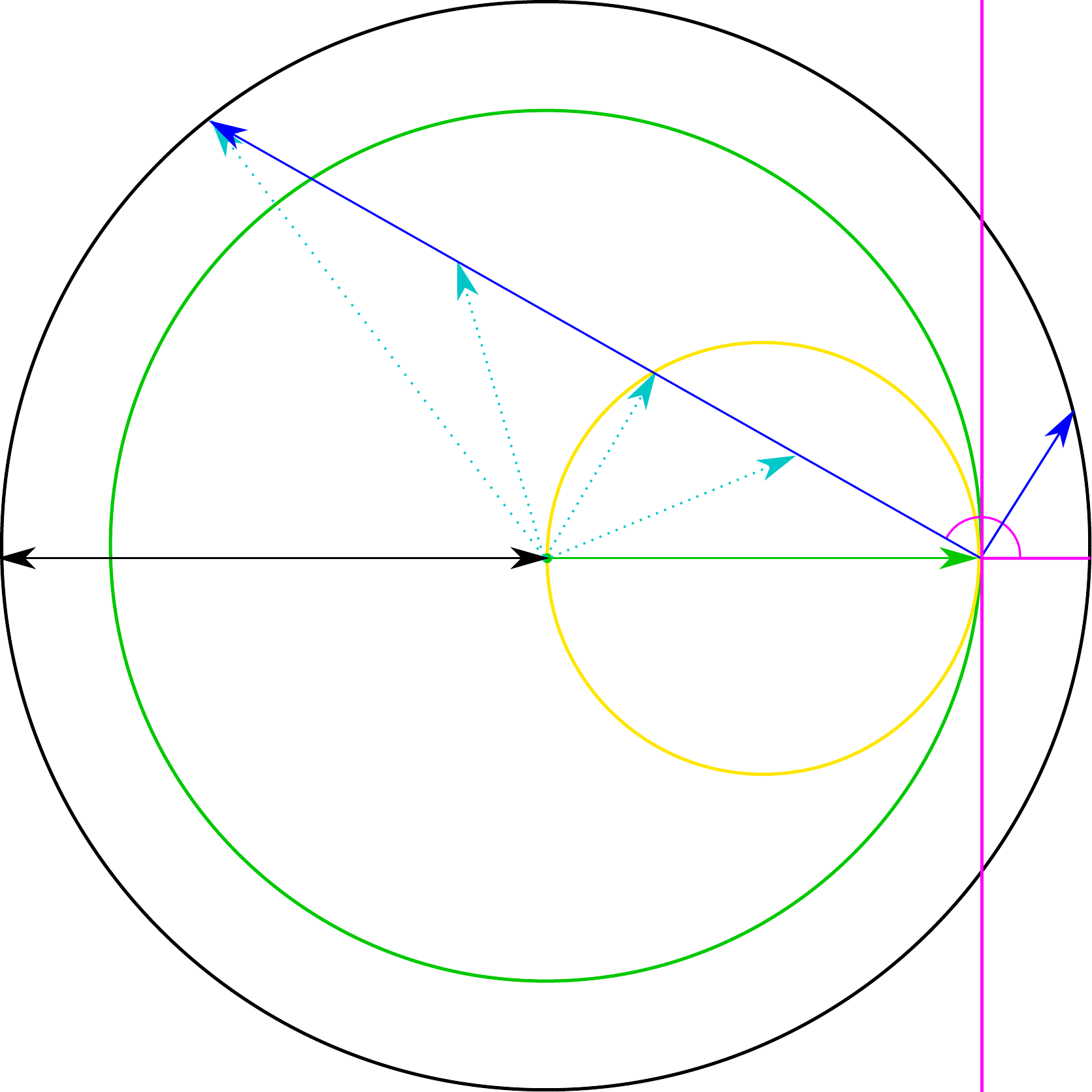}%,grid,tics=10
		\put (80.5,95) {\color{mypurple}$y<0$ \ \  $y>0$}
		\put (12,43) {neutrino sphere $R_\nu$}
		\put (69,43) {\color{mygreen} $\vec{r}$}
		\put (30,63) {\color{mylightblue} $\vec{r}\,'$}
		%\put (42,77) {\color{blue} $\vec{r}_{\theta}^{\;+}$}
		%\put (73,60) {\color{red} $\vec{r}_{\theta}^{\;-}$}
        \put (56,71) {\color{blue} $\vec{r}_{\theta}$}
		\put (85,53) {\color{mypurple}$\theta$}
		\put (95,52) {\color{blue} $\vec{r}_{\theta}$}
	\end{overpic}
\caption{The production and propagation of $Z^{\prime}$ boson inside the neutrinosphere. The magenta line divides the region with $y>0$ ($\theta< \pi /2$) and that with $y<0$ ($\theta> \pi/2$). Additionally, $r^{\prime}$ denotes the distance from the origin to the position of $Z^{\prime}$ in propagation. %The notation $r_{\theta^+}$ ($r_{\theta^-}$) means that $r^{\prime}$ increases (decreases) as $r_{\theta}$ increases.  
The vector $r_{\theta}$ denotes the trajectory of $Z'$ boson after its production. }
  \label{fig:Spherel}
\end{figure}

As the coupling strength $g_{Z'}$ increases, the re-absorption and decays of $Z^{\prime}$ bosons are not negligible. These attenuation processes impede the free-streaming of $Z'$ bosons and are considered by introducing the attenuation factor 
%\cite{Chang, Croon:2020lrf}:
\begin{equation}\label{eq.Abar}
\mathcal{A} = \frac{1}{2}\int^1_{-1}\exp\left(-\int_0^{r^{\rm max}_\theta}\Gamma_{\rm abs}dr_\theta\right)dy,
\end{equation}
where $\Gamma_{abs}$ is the reduced absorption rate \cite{Caputo:2021rux} of the $Z'$ boson. The integration over $y\equiv \cos\theta$ and the factor $1/2$ represent the averaging of $Z^{\prime}$ propagation direction inside the neutrinosphere. As shown by Fig.~\ref{fig:Spherel}, $\theta$ is the angle between the $Z^{\prime}$ propagation direction and the radial vector $\vec{r}$ from the origin to the point of $Z^{\prime}$ production. Finally, $r_\theta$ is the distance from where the $Z'$ boson is created.
%The meanings of $y$ and $r_\theta$ are illustrated by Fig.~\ref{fig:Spherel}.
%$y=\cos\theta$ and $r_\theta$ the distance from where the $Z'$ boson is created.  %$\Gamma_{abs}$ is the absorption rate of the $Z'$ boson.

%\begin{figure}[htbp]
%  \centering
%   \includegraphics[width=.7\columnwidth]{Spherel.pdf}
%  \caption{The production and propagation of $Z^{\prime}$ boson inside the neutrinosphere. The magenta line divides the region with $y>0$ ($\theta< \pi /2$) and that with $y>0$ ($\theta> \pi/2$). Additionally, $r^{\prime}$ denotes the distance from the origin to the position of $Z^{\prime}$ in propagation. The notation $r_{\theta^+}$ ($r_{\theta^-}$) means that $r^{\prime}$ increases (decreases) as $r_{\theta}$ increases.  }
%  \label{fig:Spherel}
%\end{figure}

%Implementing the attenuation factor $\mathcal{A}$, 
The $Z'$ boson luminosity with the attenuation factor implemented is given by~\cite{Chang:2016ntp, Croon:2020lrf, Caputo:2021rux, Caputo:2022rca, Bollig:2020xdr, Lucente:2020whw} 
\begin{equation}
L_{Z'} = \int^{V(R_\nu)} dV\int\frac{d^3k}{(2\pi)^3}\mathcal{A}(\omega,r,T,R_{\rm far})\omega\Gamma_{\rm prod}(\omega,T). \label{eq.LuminosityZ'}
\end{equation}
It is interesting to see that Eq.~\eqref{eq.LuminosityZ'} together with Raffelt's criterion, Eq.~\eqref{eq.RaffeltC}, unifies and addresses both the free-streaming and trapping constraints. $R_{\rm far}$ is the radius of the spherical volume for the attenuation processes, which would be taken as  $R_{\nu}$ in our study. 
%The reason will be provided in a later context.
Eq.~\ref{eq.LuminosityZ'} together with Raffelt's criterion is also known as the modified luminosity criterion in some recent literature~\cite{Caputo:2021rux, Caputo:2022rca, Lucente:2020whw}. 
Although these studies share similarities, some distinctions require clarifications.
We note that Refs.~\cite{Chang:2016ntp, Croon:2020lrf, Lucente:2020whw} simplify the calculation by 
%excluding the gravitational redshift corrections and 
considering only $+z$ and $-z$ directional averaging of $Z^{\prime}$ propagation in PNS for evaluating the attenuation factor. 
% for computational reasons. 
However, %as we shall demonstrate, 
the approximation of employing the single-directional averaging is clearly seen not justified in the large coupling regime as we shall demonstrate in Appendix E.
Furthermore, it has been demonstrated that gravitational redshift effects result in about $20\%-30\%$ corrections to the luminosity of the extra neutral boson~\cite{Caputo:2021rux}. %However, as we will soon demonstrate, employing single-directional averaging will alleviate the constraints in the large coupling regime. 
Finally, Caputo {\it et al.}~\cite{Caputo:2021rux, Caputo:2022rca} also pointed out that a reduced absorption rate should be used instead of an ordinary absorption rate in the attenuation calculation, which has not been implemented in some previous studies such as Refs.~\cite{Chang:2016ntp, Croon:2020lrf} and an early version of Ref.~\cite{Bollig:2020xdr}. As the reduced absorption rate differs from the ordinary absorption rate by a factor of $1-\exp(-\omega/T)$ in the optical depth, the $Z'$ luminosity will receive an enhancement.  
Although both gravitational redshift effects and reduced absorption rate would affect constraints on the model parameters, we have chosen to neglect these effects in the current work and focus on  
%Although this enhancement could impose more stringent constraints on the model parameters, 
%we have chosen to disregard these factors — gravitational redshift correction and reduced absorption rate — for the sake of simplicity, 
%allowing for 
the direct 
comparison with the findings of %Chang {\it et al.}
Refs.~\cite{Chang:2016ntp, Croon:2020lrf} at the trapping regime under formulae given by Eqs.~(\ref{eq.Abar}) and~(\ref{eq.LuminosityZ'}).

\section{Production, absorption, and decay of \texorpdfstring{$Z^{\prime}$}{} boson in supernovae} \label{sec.03}

In $U(1)_{L_\mu-L_\tau}$ model, $Z'$ bosons can be produced at the tree level through pair coalescence and semi-Compton processes depicted in Fig.~\ref{fig.emissionprocesses}. $Z'$ bosons can also couple to photons through an effective kinetic mixing generated by the intermediate muon or tau loops as shown by the third diagram of Fig.~\ref{fig.emissionprocesses}. Such a kinetic mixing enables the production of $Z'$ boson through the neutron-proton loop bremsstrahlung process, $n+p\to n+p+Z^{\prime}$. The production of $Z'$ bosons in supernovae is predominantly governed by these three processes, while their inverse processes then contributes to the absorption of $Z^{\prime}$~\cite{Croon:2020lrf}, which will be discussed in Sec.~\ref{sec.emissivity} and \ref{sec.attenuationlength} respectively.

\begin{comment}
    \begin{figure}[htbp]
	\begin{center}
		\includegraphics[width=.3\columnwidth]{PC.pdf}
		\includegraphics[width=.3\columnwidth]{SC.pdf} 		\includegraphics[width=.3\columnwidth]{LB.pdf} 
	\caption{Feymann Diagrams}
	\end{center}
\end{figure}
\end{comment}

\subsection{The emissivity spectrum \texorpdfstring{$d\dot{\epsilon}/d\omega$}{} of \texorpdfstring{$Z'$}{} in SN} \label{sec.emissivity}

\noindent
\begin{minipage}{1.0\columnwidth}%
	\begin{figure}[H]%
		\vspace{-1.0em}
		\begin{minipage}{.33333\columnwidth} %don't touch this unless you really know what you're doing
			\begin{figure}[H]
				\centering
				\begin{fmffile}{paircoalescence}
					\setlength{\unitlength}{.8\columnwidth}
					\vspace{1em}
					\begin{fmfgraph*}(0.9,0.5)
						%\fmfstraight
						\fmfpen{1.0}
						\fmfleft{i1,i2}
						\fmfright{o1}
						\fmf{fermion}{i2,v1}
						\fmf{fermion}{v1,i1}
						\fmf{photon,tension=1.5}{v1,o1}
						\fmfv{l=$\nu,,\mu^-$, l.a=180}{i2}
						\fmfv{l=$\bar{\nu},,\mu^+$, l.a=180}{i1}
						\fmfv{l=$Z'$, l.a=0}{o1} 
					\end{fmfgraph*}%
				\end{fmffile}%
				\caption*{Pair Coalescence}
			\end{figure}%
		\end{minipage}%
		\begin{minipage}{.33333\columnwidth} %requires mpost command
			\begin{figure}[H]
				\centering
				\begin{fmffile}{semicompton}
					\setlength{\unitlength}{.8\columnwidth}
					\vspace{1em}
					\begin{fmfgraph*}(0.9,0.5)
						\fmfpen{1.0}
						\fmfleft{i1,i2}
						\fmfright{o1,o2}
						\fmf{fermion}{i2,v2}
						\fmf{fermion}{v1,o1}
						\fmf{photon}{v2,o2}
						\fmf{photon}{i1,v1}
						\fmfv{l=$\mu^-$, l.a=180}{i2}
						\fmfv{l=$\gamma$, l.a=180}{i1}
						\fmfv{l=$Z'$, l.a=0}{o2}
						\fmfv{l=$\mu^-$, l.a=0}{o1} 
						\fmf{fermion,label=,tension=0}{v2,v1} 
					\end{fmfgraph*}%
				\end{fmffile}%
				\caption*{Semi-Compton}
			\end{figure}%
		\end{minipage}%
		\begin{minipage}{.33333\columnwidth} %requires mpost command
			\begin{figure}[H]
				\centering
				\begin{fmffile}{loopbrem}
					\setlength{\unitlength}{.8\columnwidth}
					\vspace{1em}
					\begin{fmfgraph*}(0.9,0.5)
						\fmfstraight
						\fmfpen{1.0}
						\fmfleft{i1,i2,i3}
						\fmfright{o1,o2,o3}
						\fmf{fermion}{i2,v2}
						\fmf{fermion}{v1,o1}
						\fmf{fermion}{v2,o2}
						\fmf{fermion}{i1,v1}
						\fmffreeze
						%\fmf{phantom}{o1,o3,o2}
						\fmf{phantom}{v2,v3,h1,o2}
						\fmffreeze
						\fmfkeep{fermion}
						\fmf{phantom}{v3,v4,h5,o3}
						\fmf{photon}{v3,v4}
						\fmf{photon}{v5,o3}
						\fmf{fermion,left,label=$\mu,,\tau$,tension=.6}{v4,v5}
						\fmf{fermion,left,tension=.6}{v5,v4}
						%\fmf{phantom,label=$\Delta^{++}$}{v3,o3}
						%\fmf{dashes,label=,tension=0}{v3,o3} 
						\fmfv{l=$p$, l.a=180}{i2}
						\fmfv{l=$n$, l.a=180}{i1}
						\fmfv{l=$p$, l.a=0}{o2}
						\fmfv{l=$n$, l.a=0}{o1} 
						\fmfv{l=$Z'$, l.a=0}{o3}
						\fmf{dashes,label=,tension=0}{v2,v1} 
					\end{fmfgraph*}%
				\end{fmffile}%
				\caption*{Loop Brem}
			\end{figure}%
		\end{minipage}%
		\caption{Dominant $Z'$ emission processes in a core-collapse supernova.}%
		\label{fig.emissionprocesses}%
	\end{figure}%
\end{minipage}\vspace{12pt plus 2pt minus 2pt}%

At the energy scale of a supernova, the main $Z'$ production processes are pair coalescence, semi-Compton, and loop bremsstrahlung. Fig.~\ref{fig.emissionprocesses} displays the corresponding Feynman diagrams. We calculate the emissivity spectrum for each production process in this section. The relation between emissivity and production rate is
\begin{align}
	d\dot{\epsilon} &= \frac{d^3k}{(2\pi)^3}\sum_{i=1}^3 \omega \Gamma^i_{\rm prod} = \frac{\omega^3 }{2\pi^2} \sqrt{1-\frac{m_{Z'}^2}{\omega^2}} \sum_{i=1}^3\Gamma^i_{\rm prod} d\omega.
\end{align}
%\begin{align}
%	d\dot{\epsilon} &= \frac{d^3k}{(2\pi)^3}\omega \Gamma \nonumber\\
%	&= \frac{k^2dk}{2\pi^2}\omega\Gamma \qquad (\text{note that } k^2dk = \omega^2\sqrt{1-\frac{m_{Z'}^2}{\omega^2}} d\omega) \nonumber\\
%	&= \frac{\omega^3 }{2\pi^2} \sqrt{1-\frac{m_{Z'}^2}{\omega^2}} \Gamma d\omega.
%\end{align}
The emissivity spectrum $d\dot{\epsilon}/d\omega$ including all production rates $\Gamma^{i}_{\text{prod.}}$ in Fig.~\ref{fig.emissionprocesses} for $i\in$ \{pair-coalescence, semi-Compton, loop bremsstrahlung\} is
\begin{equation}
	\frac{d\dot{\epsilon}}{d\omega} = \frac{\omega^3 }{2\pi^2} \sqrt{1-\frac{m_{Z'}^2}{\omega^2}} \sum_{i=1}^{3}\Gamma^{i}_{\text{prod}}  \label{eq.emisspectrum},
\end{equation}
in which the production rate of each process $\Gamma^{i}_{\text{prod}}$ can be obtained by integrating the general phase space formula
\begin{align}\label{eq.prod_formula}
	d\Gamma &= \frac{1}{2\omega} \prod_{i} \frac{d^3 p_i f(p_i)}{(2\pi)^3 2E_i} \prod_{f} \frac{d^3 p_f(1\pm f(p_f))}{(2\pi)^3 2E_f} (2\pi)^4\delta^4(\sum_{i} p_i - \sum_{f}p_f -k) |\mathcal{M}|^2,
\end{align}
where the indices $i$ and $f$ refer to the initial- and final-state particles, respectively, apart from the new boson $Z'$. The function $f(p)$ represents the thermal statistical distribution of each individual particle. The use of a plus or minus sign depends on if the particle adheres to Fermi-Dirac or Bose-Einstein statistics. For a fermion, a Pauli blocking factor $(1-f(p))$ should be applied. On the other hand, for a boson, a Bose enhancement factor of $(1+f(p))$ should be used. The detailed derivation is provided in Appendices~\ref{sec.Decay_pC}, \ref{sec.SemiComp} and \ref{sec.LoopBrem} for pair-coalescence, semi-Compton and loop bremsstrahlung processes shown in Fig.~ \ref{fig.emissionprocesses}, respectively. 

After performing the phase space integral in Eq.~(\ref{eq.prod_formula}), the $Z'$ production rate $\Gamma_{\mu^- + \mu^+ \rightarrow Z'}$ by muon pair coalescence (the first diagram in Fig.~\ref{fig.emissionprocesses}) is given by (see Appendix \ref{sec.Decay_pC})
\begin{equation}
		\Gamma_{\mu^- + \mu^+ \rightarrow Z'} \approx \frac{g_{Z'}^2 m_{Z'}^2}{4\pi\omega} \left( 1+\frac{2m_{\mu}^2}{m_{Z'}^2} \right) \sqrt{1-\frac{4m_{\mu}^2}{m_{Z'}^2}}\times e^{-\omega/T}.
	\label{eq.pairCoalmu}
\end{equation}
Similarly, the $Z'$ production rate $\Gamma_{\nu + \bar{\nu} \rightarrow Z'}$ via neutrino (including both $\nu_\mu$ and $\nu_\tau$) pair coalescence can be calculated in the same way. It is given by
\begin{equation}
	\Gamma_{\nu + \bar{\nu} \rightarrow Z'} \approx \frac{g_{Z'}^2 m_{Z'}^2}{4\pi \omega} \times e^{-\omega/T}.  \label{eq.pairCoalnu}
\end{equation}
We note that the allowed polarizations for muons and neutrinos are different. Unlike muons having two polarizations, only left-handed neutrinos can contribute to the matrix element squared $|\mathcal{M}|^2$. The equations \eqref{eq.pairCoalmu} and \eqref{eq.pairCoalnu} share the same prefactor as both $\nu_\mu$ and $\nu_\tau$ have been accounted for.

For the semi-Compton process (the second diagram in Fig.~\ref{fig.emissionprocesses}), the production rate $\Gamma_{\text{sC}}$ is given by (see Appendix \ref{sec.SemiComp})
\begin{equation} \label{eq.sCproduction}
		\Gamma_{\text{sC}}	\approx \frac{8\pi\alpha\alpha_{Z'}}{3m_\mu^2}\frac{n_{\mu}F_{\text{deg}}}{e^{\omega/T}-1}\sqrt{1-\frac{m_{Z'}^2}{\omega^2}}, 
\end{equation}
where $\alpha_{Z'}\equiv g_{Z'}/4\pi$, $n_{\mu}$ is the muon number density, and the degeneracy factor $F_{\text{deg}}\in[0,1]$ is considered as a simple Pauli-blocking factor correction to the phase space of the final-state muon. The definition and determination of $F_{\text{deg}}$ are discussed in Appendix~\ref{sec.SemiComp}. 

Finally, since the loop bremsstrahlung process (the last diagram in Fig.~\ref{fig.emissionprocesses}) is considerably more intricate than the other processes, the emissivity spectrum is provided directly instead of the production rate:
%\begin{equation}
%		\frac{d\dot{\epsilon}_{\text{LB}}}{d\omega} = \frac{2n_n n_p \alpha\epsilon^2 }{3(\pi M T)^{3/2}}  \sqrt{1-\left(\frac{m_d}{\omega}\right)^2} \left[ 2+\left(\frac{m_d}{\omega}\right)^2 \right] \int_{\omega}^{\infty} dT_{\text{cm}} e^{-T_{\text{cm}}/T} T_{\text{cm}}^2 \sigma^{(2)}_{np}(T_{\text{cm}}).
%\end{equation}
\begin{equation}
		\frac{d\dot{\epsilon}_{\text{LB}}}{d\omega} = \frac{2n_n n_p \alpha\epsilon^2 }{3(\pi M T)^{3/2}}  \sqrt{1-\left(\frac{m_{Z'}}{\omega}\right)^2} \left[ 2+\left(\frac{m_{Z'}}{\omega}\right)^2 \right] \int_{\omega}^{\infty} dT_{\text{cm}} e^{-T_{\text{cm}}/T} T_{\text{cm}}^2 \sigma^{(2)}_{np}(T_{\text{cm}}), 
\end{equation}
where $T_{\rm cm}$ is the kinetic energy of the initial particles in the center-of-momentum frame, $M$ is the nucleon mass, $n_n$ and $n_p$ are number densities of neutrons and protons, respectively, $\epsilon$ is an effective $\gamma-Z^{\prime}$ mixing $\epsilon Z^{\prime}_{\mu\nu}Z^{\mu\nu}/2$ induced by one-loop diagrams. It is related to $g_{Z'}$ and lepton masses by
\begin{equation}
\epsilon=-\frac{eg_{Z'}}{2\pi^2}\int_0^1 x(1-x) \ln \left[\frac{m_\tau^2-x(1-x)k^2}{m_\mu^2-x(1-x)k^2}\right],  
\label{induced-mixing}
\end{equation}
where $k=(\omega,\vec{k})$ is the 4-momentum of final-state $Z^{\prime}$ boson with $k^2=m_{Z'}^2$.
The definition of $\sigma^{(2)}_{np}(T_{\text{cm}})$ in the integrand is
\begin{align}
	\sigma^{(2)}_{np}(T_{\text{cm}}) &\equiv \int_{-1}^{1} (1-\cos\theta_{\text{cm}})\frac{d\sigma_{np}}{d\cos\theta_{\text{cm}}} d\cos\theta_{\text{cm}},
\end{align}
where $d\sigma_{np}/d\cos\theta_{\text{cm}}$ is the neutron-proton differential cross-section that can be extracted from experiments. The derivation of these formulae is given in Appendix \ref{sec.LoopBrem}.

\begin{comment}
The production rate for neutrino-pair coalescence is determined by
\begin{equation}
\Gamma_{\rm PC}(\omega) = \frac{1}{2\omega}\int\frac{d^3p_1}{2E_1(2\pi)^3}\frac{d^3p_2}{2E_2(2\pi)^3}(2\pi)^4|\mathcal{M}_{\nu\bar{\nu}\rightarrow Z'}|^2f_\nu(E_1)f_\nu(E_2)\delta^4(p_1+p_2-p_{Z'}),
\end{equation}
where $f_\nu(E_i)=(\exp(E^i/T)+1)^{-1}$ denotes the neutrino thermal distribution at temperature $T$ and $E_1=E_2=\omega/2$ in the $Z'$ rest frame. Since $|\mathcal{M}_{\nu\bar{\nu}\rightarrow Z'}|^2=|\mathcal{M}_{Z'\rightarrow\nu\bar{\nu}}|^2$, one has
\begin{equation}
\Gamma_{\rm PC}(\omega) = \Gamma_{dec}\left(\frac{1}{\exp(\omega/2T)+1}\right)^2,
\end{equation}
with $Z'$ decay rate $\Gamma_{dec}=\alpha_{Z'}m_{Z'}/3$. Hence, the production rate for neutrino-pair coalescence is given by
\begin{equation}
\Gamma_{\rm PC}(\omega) = \frac{\alpha_{Z'}m^2_{Z'}}{3\omega}\left(\frac{1}{\exp(\omega/2T)+1}\right)^2
\end{equation}
The inverse process of the neutrino-pair coalescence is just $Z'$ decay into neutrino pair such that the corresponding absorption rate is just the $Z'$ decay rate
\end{comment}

\begin{comment}
\subsection{Semi-Compton scattering}

The production rate for muon semi-Compton scattering is determined by

\begin{equation}
\Gamma_{\rm SC}(\omega) = \frac{1}{2\omega}\int\frac{d^3p_1}{2E_1(2\pi)^3}\frac{d^3p_2}{2E_2(2\pi)^3}\frac{d^3q}{2\omega_q(2\pi)^3}(2\pi)^4|\mathcal{M}_{\mu^-\gamma\rightarrow\mu^- Z'}|^2f_\mu(E_1)(1-f_\mu(E_2))f_\gamma(\omega_q)\delta^4(p_1+p_2-p_{Z'}),
\end{equation}
\end{comment}

\subsection{The attenuation length of \texorpdfstring{$Z'$}{} in SN}\label{sec.attenuationlength}

\noindent
\begin{minipage}{1.0\columnwidth}%
	\begin{figure}[H]%
		\vspace{-1.0em}
		\begin{minipage}{.33333\columnwidth} %don't touch this unless you really know what you're doing
			\begin{figure}[H]
				\centering
				\begin{fmffile}{decay}
					\setlength{\unitlength}{.8\columnwidth}
					\vspace{1em}
					\begin{fmfgraph*}(0.9,0.5)
						%\fmfstraight
						\fmfpen{1.0}
						\fmfleft{i1}
						\fmfright{o1,o2}
						\fmf{fermion}{o2,v1}
						\fmf{fermion}{v1,o1}
						\fmf{photon,tension=1.5}{i1,v1}
						\fmfv{l=$\nu,,\mu^-$, l.a=0}{o2}
						\fmfv{l=$\bar{\nu},,\mu^+$, l.a=0}{o1}
						\fmfv{l=$Z'$, l.a=180}{i1} 
					\end{fmfgraph*}%
				\end{fmffile}%
				\caption*{$Z'$ Decay}
			\end{figure}%
		\end{minipage}%
		\begin{minipage}{.33333\columnwidth} %requires mpost command
			\begin{figure}[H]
				\centering
				\begin{fmffile}{invsemicompton}
					\setlength{\unitlength}{.8\columnwidth}
					\vspace{1em}
					\begin{fmfgraph*}(0.9,0.5)
						\fmfpen{1.0}
						\fmfleft{i1,i2}
						\fmfright{o1,o2}
						\fmf{fermion}{i1,v1}
						\fmf{fermion}{v2,o2}
						\fmf{photon}{v1,o1}
						\fmf{photon}{i2,v2}
						\fmfv{l=$\mu^-$, l.a=180}{i1}
						\fmfv{l=$\gamma$, l.a=0}{o1}
						\fmfv{l=$Z'$, l.a=180}{i2}
						\fmfv{l=$\mu^-$, l.a=0}{o2} 
						\fmf{fermion,label=,tension=0}{v1,v2} 
					\end{fmfgraph*}%
				\end{fmffile}%
				\caption*{Inv Semi-Comp}
			\end{figure}%
		\end{minipage}%
		\begin{minipage}{.33333\columnwidth} %requires mpost command
			\begin{figure}[H]
				\centering
				\begin{fmffile}{invloopbrem}
					\setlength{\unitlength}{.8\columnwidth}
					\vspace{1em}
					\begin{fmfgraph*}(0.9,0.5)
						\fmfstraight
						\fmfpen{1.0}
						\fmfleft{i1,i2,i3}
						\fmfright{o1,o2,o3}
						\fmf{fermion}{i2,v2}
						\fmf{fermion}{v1,o1}
						\fmf{fermion}{v2,o2}
						\fmf{fermion}{i1,v1}
						\fmffreeze
						%\fmf{phantom}{o1,o3,o2}
						\fmf{phantom}{v2,v3,h1,i2}
						\fmffreeze
						\fmfkeep{fermion}
						\fmf{phantom}{v3,v4,h5,i3}
						\fmf{photon}{v3,v4}
						\fmf{photon}{v5,i3}
						\fmf{fermion,left,tension=.6}{v4,v5}
						\fmf{fermion,left,label=$\mu,,\tau$,tension=.6}{v5,v4}
						%\fmf{phantom,label=$\Delta^{++}$}{v3,o3}
						%\fmf{dashes,label=,tension=0}{v3,o3} 
						\fmfv{l=$p$, l.a=180}{i2}
						\fmfv{l=$n$, l.a=180}{i1}
						\fmfv{l=$p$, l.a=0}{o2}
						\fmfv{l=$n$, l.a=0}{o1} 
						\fmfv{l=$Z'$, l.a=180}{i3}
						\fmf{dashes,label=,tension=0}{v2,v1} 
					\end{fmfgraph*}%
				\end{fmffile}%
				\caption*{Inv Loop Brem}
			\end{figure}%
		\end{minipage}%
		\caption{Dominant $Z'$ reabsorption processes in core-collapse supernova.}%
		\label{fig.reabsorptionprocesses}%
	\end{figure}%
\end{minipage}\vspace{12pt plus 2pt minus 2pt}%

Here we discuss the attenuation length $\lambda_{\text{att}}$, or in other words, the mean free path of $Z'$. It describes the survival distance of the $Z'$ boson that is produced by either of the aforementioned production processes. The dominant decay and reabsorption processes are illustrated in Fig.~\ref{fig.reabsorptionprocesses}. They are considered as the inverse processes of those given in Fig.~\ref{fig.emissionprocesses}. In this section, we shall compute the inverse of the attenuation length, denoted by $1/\lambda_{\text{att}}$. Its relation to the absorption rate $\Gamma$ (in the rest frame of $Z'$) is given by
\begin{align}
	\frac{1}{\lambda_{\text{att}}} &= \frac{\Gamma^{\text{lab}}}{v} = \Gamma\times\frac{m_{Z'}}{\omega}\bigg/ \sqrt{1-\left( \frac{m_{Z'}}{\omega} \right)^2} \label{eq.lambdatt},
\end{align}
where $\Gamma^{\text{lab}}$ and $v$ denote the absorption rate in the lab frame and the velocity of the new boson $Z'$, respectively, $m_{Z'}/\omega$ is just the inverse of $\gamma$ factor accounting for the time dilation effect in a moving frame. The last term $\sqrt{1-(m_{Z'}/\omega)^2}$ in the denominator is the speed factor $v$. Hence the calculation of $1/\lambda_{\text{att}}$ amounts to computing the absorption rate $\Gamma$ for each process. 

The absorption rate can be calculated by replacing the 4-momentum $k\rightarrow -k$ in the production rate formula Eq.~\eqref{eq.prod_formula} and the corresponding matrix element squared for absorption processes in Fig.~\ref{fig.reabsorptionprocesses}. It gives
\begin{align}\label{eq.abs_formula}
	d\Gamma &= \frac{1}{2\omega} \prod_{i} \frac{d^3 p_i f(p_i)}{(2\pi)^3 2E_i} \prod_{f} \frac{d^3 p_f(1\pm f(p_f))}{(2\pi)^3 2E_f} (2\pi)^4\delta^4(\sum_{i} p_i - \sum_{f}p_f + k) |\mathcal{M}|^2,
\end{align}
where the indices $i$ and $f$ stand for initial and final state particles. The function $f(p)$ is the thermal statistical distribution of the initial and final state particles which has already been introduced in the previous section. The detail of the derivation is provided in Appendices \ref{sec.Decay_pC}, \ref{sec.SemiComp} and \ref{sec.LoopBrem} for $Z'$ decay, semi-Compton absorption and loop bremsstrahlung absorption processes in Fig.~\ref{fig.reabsorptionprocesses}, respectively. 

The decay rate of $Z'$ through $Z'\rightarrow \mu^- + \mu^+$ is given by Eq.~\eqref{aeq.labframedecayrate_muon} (see Appendix \ref{sec.Decay_pC}):
\begin{equation}
		\Gamma_{Z'\rightarrow \mu^- + \mu^+}^{\text{lab}}
		\approx \frac{g_{Z'}^2 m_{Z'}^2}{12\pi\omega} \left( 1+\frac{2m_{\mu}^2}{m_{Z'}^2} \right) \sqrt{1-\frac{4m_{\mu}^2}{m_{Z'}^2}} 
	\label{eq.labframedecayrate_muon}.
\end{equation}
The inverse of the attenuation length $1/\lambda$ can then be obtained by dividing Eq.~\eqref{eq.labframedecayrate_muon}by a speed factor $v$ according to the relation in Eq.~\eqref{eq.lambdatt}. This gives
\begin{equation}
	\frac{1}{\lambda_{Z'\rightarrow \mu^- + \mu^+}}
	\approx \frac{g_{Z'}^2 m_{Z'}^2}{12\pi\omega} \left( 1+\frac{2m_{\mu}^2}{m_{Z'}^2} \right) \sqrt{1-\frac{4m_{\mu}^2}{m_{Z'}^2}} \bigg/ \sqrt{1-\left( \frac{m_{Z'}}{\omega} \right)^2} 
	\label{eq.att_muon}.
\end{equation}

The derivation of the decay rate for $Z' \rightarrow \nu + \bar{\nu}$ is very similar to that of $Z' \rightarrow \mu^- + \mu^+$. To obtain the decay rate $\Gamma_{Z' \rightarrow \nu + \bar{\nu}}$, one can simply take $m_\mu \rightarrow 0$ in Eq.~\eqref{eq.labframedecayrate_muon}, which gives
\begin{equation}
		\Gamma_{Z'\rightarrow \nu + \bar{\nu}}^{\text{lab}}
		\approx \frac{g_{Z'}^2 m_{Z'}^2}{12\pi\omega}
	\label{eq.labframedecayrate_neutrino}.
\end{equation}
It is worth noting that only left-handed neutrinos and right-handed anti-neutrinos can interact with the new gauge boson $Z'$. This is because, unlike muons which have two polarization states, neutrinos have only one. As a result, we have $\Gamma^{\text{lab}}_{Z'\rightarrow \nu_{\mu} + \bar{\nu}_{\mu}} = \frac{1}{2}\Gamma^{\text{lab}}_{Z'\rightarrow \mu^- +\mu^+}(m_{\mu}\rightarrow 0)$. The pre-factor in Eq.~\ref{eq.labframedecayrate_neutrino} is equal to that in Eq.~\ref{eq.labframedecayrate_muon} because both $Z'\rightarrow \nu_{\mu} + \bar{\nu}_{\mu}$ and $Z'\rightarrow \nu_{\tau} + \bar{\nu}_{\tau}$ have been taken into account (for more details, see Appendix \ref{sec.Decay_pC}). We obtain the inverse of the attenuation length $1/\lambda$ by dividing Eq.~\eqref{eq.labframedecayrate_neutrino} by the $Z'$ speed $v$, which gives
\begin{equation}
		\frac{1}{\lambda_{Z'\rightarrow \nu + \bar{\nu}}}
		\approx \frac{g_{Z'}^2 m_{Z'}^2}{12\pi\omega} \bigg/ \sqrt{1-\left( \frac{m_{Z'}}{\omega} \right)^2}
	\label{eq.att_neutrino}.
\end{equation}

The $Z'$ absorption rate through semi-Compton process (the second diagram in Fig.~\ref{fig.reabsorptionprocesses}) can be obtained by using the principle of detailed balance:
\begin{align}
	\Gamma_{\text{sC}}^{\text{abs}} &= \frac{2}{3} e^{\omega/T}\times\Gamma_{\text{sC}}^{\text{prod}}.
\end{align}
Substituting Eq.~\eqref{eq.sCproduction} into the above equation gives
\begin{align}
	\Gamma_{\text{sC}}^{\text{abs}} &\approx \frac{2}{3} e^{\omega/T} \times \frac{8\pi\alpha\alpha_{Z'}}{3m_\mu^2}\frac{n_{\mu}F_{\text{deg}}}{e^{\omega/T}-1}\sqrt{1-\frac{m_{Z'}^2}{\omega^2}} \nonumber\\
	&= \frac{ 16\pi \alpha\alpha_{Z'}}{9m_\mu^2}(1+ \frac{1}{e^{\omega/T}-1})  n_{\mu}F_{\text{deg}}  \sqrt{1-\frac{m_{Z'}^2}{\omega^2}}.
\end{align}
A more comprehensive derivation, with or without using the principle of detailed balance, is given in Appendix \ref{sec.SemiComp}. By substituting the aforementioned equation into Eq.~\eqref{eq.lambdatt}, one  obtains $1/\lambda_{\text{sC}}$ given by
\begin{equation}
	\frac{1}{\lambda_{\text{sC}}} \approx \frac{ 16\pi \alpha\alpha_{Z'}}{9m_\mu^2}(1+ \frac{1}{e^{\omega/T}-1})  n_{\mu}F_{\text{deg}} .
\end{equation}

Finally, the inverse of the attenuation length $1/\lambda$ for the $Z'$ absorption through loop-bremsstrahlung is significantly more complex than the others. A detailed derivation of this can be found in Appendix \ref{sec.LoopBrem}. The result is
\begin{equation}
		\frac{1}{\lambda_{\text{LB}}} = \frac{8n_n n_p \alpha\epsilon^2 }{9\pi\omega^3 } \left( \frac{\pi T}{M}\right)^{3/2} \left( \frac{2+(m_{Z'}/\omega)^2}{ \sqrt{1-(m_{Z'}/\omega)^2} } \right)  \frac{1}{2}\int_0^\infty dx e^{-x} x^2  \sigma^{(2)}_{np}(x).
\end{equation}

\section{Results and discussion} \label{sec.04}

\begin{figure}[thbp]
		\centering
		\includegraphics[width=.5\columnwidth]{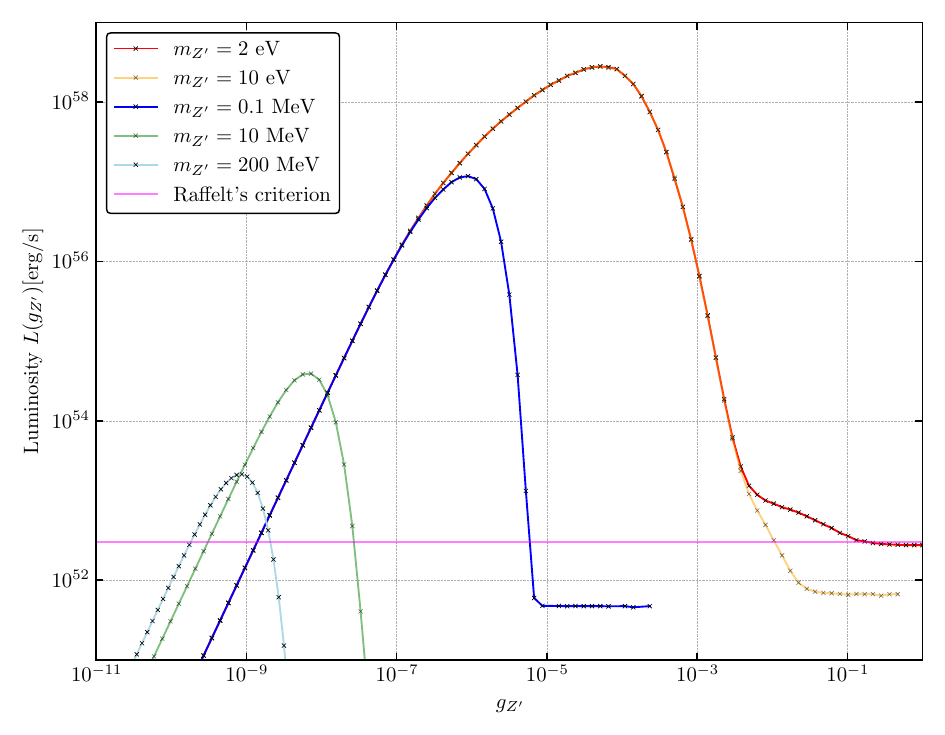}%
		\includegraphics[width=.5\columnwidth]{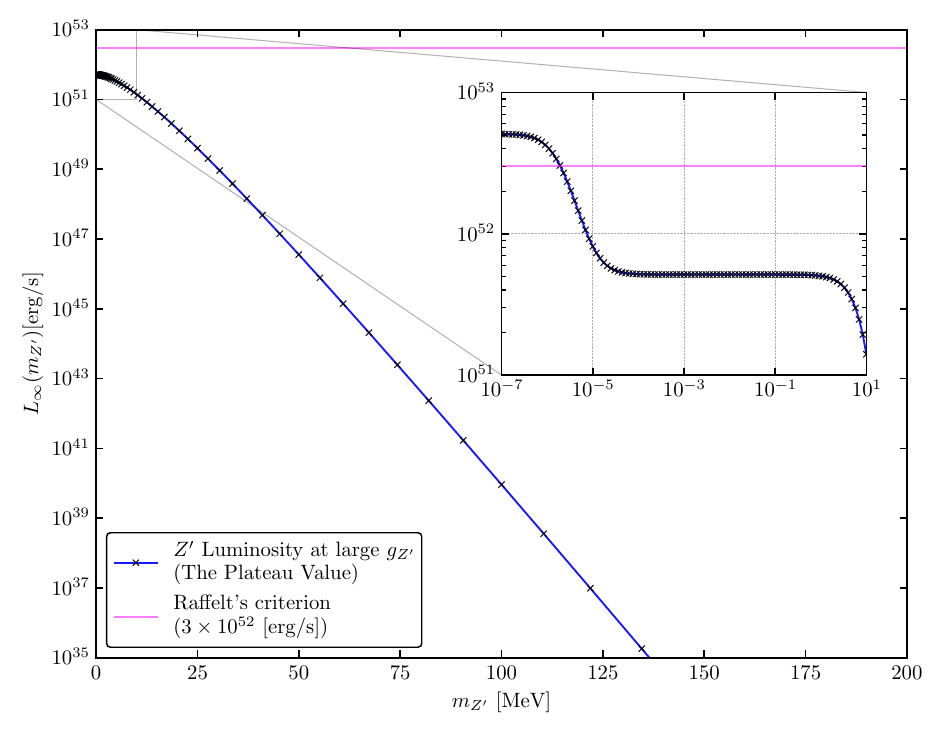}%
		\caption{(Left) Luminosity of the new gauge boson $Z'$ in various mass $m_Z'$. (Right) The plateau value $L_{\infty}$ as a function of $m_{Z'}$. The detailed behaviour of $L_{\infty}$ for $m_{Z'}$ below $10$ MeV is given by the inset.}
		\label{fig:Luminosity}
	\end{figure}

The $U(1)_{L_\mu-L_\tau}$ model introduces a new gauge boson $Z'$ that can be produced through interactions between standard model particles and unstable heavy leptons, specifically muons and taus. This is described by the interaction Lagrangian in Eq.~\eqref{Lmu-LtauLagrangian}. In this section, we present $Z'$ luminosity based upon a SN simulation SFHo18.8 by Bollig {\it et al.})~\cite{Bollig:2020xdr}. Choosing SFHo18.8 simulation enables the comparison of our calculations with those in~\cite{Croon:2020lrf}.  We account for the production and reabsorption processes discussed in the previous sections and obtain the $Z'$ luminosity through numerical calculation of Eq.\eqref{eq.LuminosityZ'} among different values of $m_{Z'}$ and $g_{Z'}$. The left panel of Fig.~\ref{fig:Luminosity} shows the $Z'$ luminosity for $m_{Z'}=\{2 \text{ eV}, 10 \text{ eV}, 0.1 \text{ MeV}, 10 \text{ MeV}, 200 \text{ MeV} \}$ as a function of the coupling $g_{Z'}$. The Raffelt bound (see Eq.~\eqref{eq.RaffeltC}) is indicated by a magenta line. 
	
Many interesting features can be seen in Fig.~\ref{fig:Luminosity} as pointed out by Croon {\it et al.}~\cite{Croon:2020lrf}. The peak value of $L_{Z'}$ goes down as $m_{Z'}$ increases is a clear evidence of the Boltzmann suppression. %The mass-dependent and mass-independent regimes of $L_{Z'}$ can be seen from the free streaming region. 
In the free streaming region, $L_{Z'}$ steadily increases with $g_{Z'}$.
On the other hand, the exponential suppression of $L_{Z'}$ can be observed from the trapping region for a sufficiently large $g_{Z'}$. The only difference between the result of Ref.~\cite{Croon:2020lrf} and that of ours is that the former does not manifest the plateau phenomenon when taking %$R_{a}\rightarrow R_{e}$ 
 $R_{\text{far}}\rightarrow R_{\nu}$ with a large $g_{Z'}$. To be precise, we observe that each luminosity curve in the left panel of Fig.~\ref{fig:Luminosity} approaches to a constant plateau value for a sufficiently large $g_{Z'}$.  
 	
 	\begin{figure}[bthp]
 		\centering
 		\includegraphics{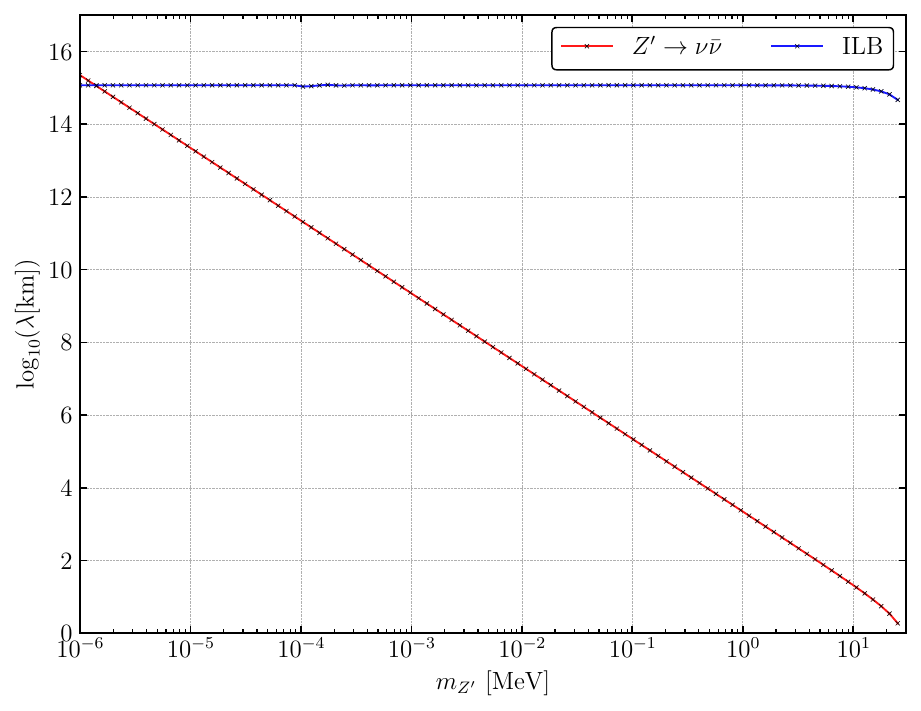}%
 		\caption{The mean free paths of $Z'\rightarrow \nu + \bar{\nu}$ decay and the inverse loop bremsstrahlung reabsorption at $\omega=30$ MeV.}
 		\label{fig:mfp}
 	\end{figure}
 	
 	%On the other hand, although we do not delve into detail here, 
 
 This non-vanishing luminosity is however not seen from Fig.~2 of Ref.~\cite{Croon:2020lrf} where the limit  $R_{\text{far}}\rightarrow R_{\nu}$ is taken for $L_{Z'}$ with $m_{Z'}=0.1$ eV. Before we provide arguments for this plateau phenomenon, we reiterates that the production and attenuation regions of $Z'$ boson should be taken as identical for consistency.  
 %The authors highlight that as taking $R_{a}=R_{e}=R_{\nu}$, more energy seems to escape from the supernova. Therefore, as a conservative approach, they use $R_{a}=100$ km (or $R_{\text{far}}=100$ km in their notation) throughout their study. However, this energy can potentially be deposited back into the supernova, which could affect the neutrino luminosity, shockwave formation and revival, as well as the evolution of the proto-neutron star.
%To demonstrate that this choice underestimates $L_{Z'}$, let's consider the meanings of $R_{a}$ and $R_{e}$. $R_{a}$ is the upper limit of the integration of the attenuation factor $\mathcal{A}$, which determines the volume of the region contributing to supernova cooling. In contrast, $R_{e}$ is the upper limit of the integration of the emissivity, which determines the volume of the region where $Z'$ production contributes to the luminosity $L_{Z'}$. Thus, using different values of $R_{a}$ and $R_{e}$ implies different supernova volumes for attenuation and emissivity calculations. 
On the other hand, Croon {\it et al.}~\cite{Croon:2020lrf} present their results by taking $R_{\rm far}\gg R_{\nu}$. 
%it is important to acknowledge that $Z'$ is still being produced within the region enclosed by $R_{a}$ and $R_{e}$. It would be unfair to favour attenuation by taking a greater upper limit of integration. Thus, to ensure a fair treatment of $R_{a}$ and $R_{e}$, we use $R_{e}=R_{a}=R_{\nu}$ exclusively in this work.
	
%Figure~\ref{fig:Luminosity} shows that the $Z'$ luminosity reaches a constant plateau value as the coupling $g_{Z'}$ increases. 
The above-mentioned plateau phenomenon can be understood by a simple argument. In the large $g_{Z'}$ limit, only those $Z'$ bosons produced within one interaction length from the surface of the neutrinosphere can exit the sphere and contribute to $L_{Z'}$. The volume of this spherical shell is $4\pi R_{\nu}^2\lambda$ with $\lambda$ the interaction length. The number of $Z^{\prime}$ bosons produced in this volume is proportional to $Z^{\prime}$ production probability multiplied by $4\pi R_{\nu}^2\lambda$. With the former scaled as $g_{Z'}^2$ and the latter as $g_{Z'}^{-2}$, the overall contribution to $L_{Z'}$ is then independent of $g_{Z'}$. The details for calculating this limiting luminosity, denoted by $L_{\infty}$, are given in Fig.~\ref{fig.Sphere_e} and Eq.~\eqref{eq.Linf}. The expected plateau value $L_{\infty}$ as a function of $m_{Z'}$ is given in the right panel of Fig.~\ref{fig:Luminosity}.

At this juncture, we like to stress that the $Z^{\prime}$ luminosity in the large coupling regime cannot be accurately reproduced using the Stefan-Boltzmann approximation (see \cite{Chang:2016ntp,DeRocco:2019jti,Caputo:2021rux}). Since the emission and reabsorption rates depend significantly on the $Z'$ energy $\omega$, the decoupling radius also varies. Consequently, the luminosity value obtained from Eq.~(\ref{eq.LuminosityZ'}) will not agree with the Stefan-Boltzmann approximation. To explicitly illustrate the above-mentioned energy dependencies, let us consider the small $m_{Z^{\prime}}$ limit. We note that the dominant reabsorption channels near the neutrinosphere for small $m_{Z'}$ are $Z'$ decay:
 	\begin{align}
 		\Gamma_{Z' \rightarrow \nu + \bar{\nu}} &\propto m_{Z'}^2 \omega^{-1}
         \label{eq:decay_scaling}	
  \end{align}
 	and inverse loop bremsstrahlung $n+p+Z'\to n+p$:
 	\begin{align}
 		\Gamma_{\rm ILB} &\propto \omega^{-3}.
       \label{eq:LB_scaling}	
  \end{align}
  The inverse semi-Compton process is not considered here since muon number density is negligible near the neutrinosphere.  
 
 Both the decay length of $Z'\rightarrow \nu + \bar{\nu}$ and the interaction length of inverse loop bremsstrahlung at $\omega = 30$ MeV are shown in Fig.~\ref{fig:mfp}. Using Eqs.~(\ref{eq:decay_scaling}) and
 (\ref{eq:LB_scaling}) along with the information in Fig.~\ref{fig:mfp}, one can argue through simple scaling that the inverse loop bremsstrahlung reabsorption dominates the $Z^{\prime}$ decay when $\omega m_{Z'} < 21 \, {\rm keV}^2$. Considering that the temperature at the surface of the neutrinosphere is $\sim 3 \, {\rm MeV}$, we can expect $\omega \sim {\rm MeV}$ or softer. Consequently, it can be deduced that inverse loop bremsstrahlung reabsorption dominates when $m_{Z'} < 2.1\cdot 10^{-2}  \, {\rm keV}$. 
 On the contrary, $Z'\rightarrow \nu + \bar{\nu}$ dominates inverse loop bremsstrahlung if $\omega m_{Z'} > 21 \, {\rm keV}^2$. Since $\omega \geq m_{Z'}$, the condition $\omega m_{Z'} > 21 \, {\rm keV}^2$ automatically holds if $m_{Z'}> 4.6$ keV. For $2.1\cdot 10^{-2} \leq (m_{Z'}/{\rm keV})\leq 4.6$,  the $Z'$ energy $\omega$ determines which channel is more important. 
 
 Given the strong dependence of the inverse loop bremsstrahlung event rate on $\omega^{-3}$, it is clearly impossible to utilize a single decoupling radius to approximate the $Z'$ luminosity in the trapping regime.
 Thus, the actual $Z'$ luminosity cannot be accurately estimated by the Stefan-Boltzmann law, as indicated in~\cite{Caputo:2021rux} (particularly at Section III.B.7). In fact, for the dark photon case, it has been pointed out in~\cite{Chang:2016ntp} that the dark photon luminosity from PNS in the trapping regime is significantly greater than that predicted by the Stefan-Boltzmann approximation. Therefore, in the current case the actual value of $Z^{\prime}$ must be obtained from a full calculation by Eq.~(\ref{eq.LuminosityZ'}) with specific supernova simulation data.

\begin{figure}[thbp]
		\centering
		\includegraphics[width=.5\columnwidth]{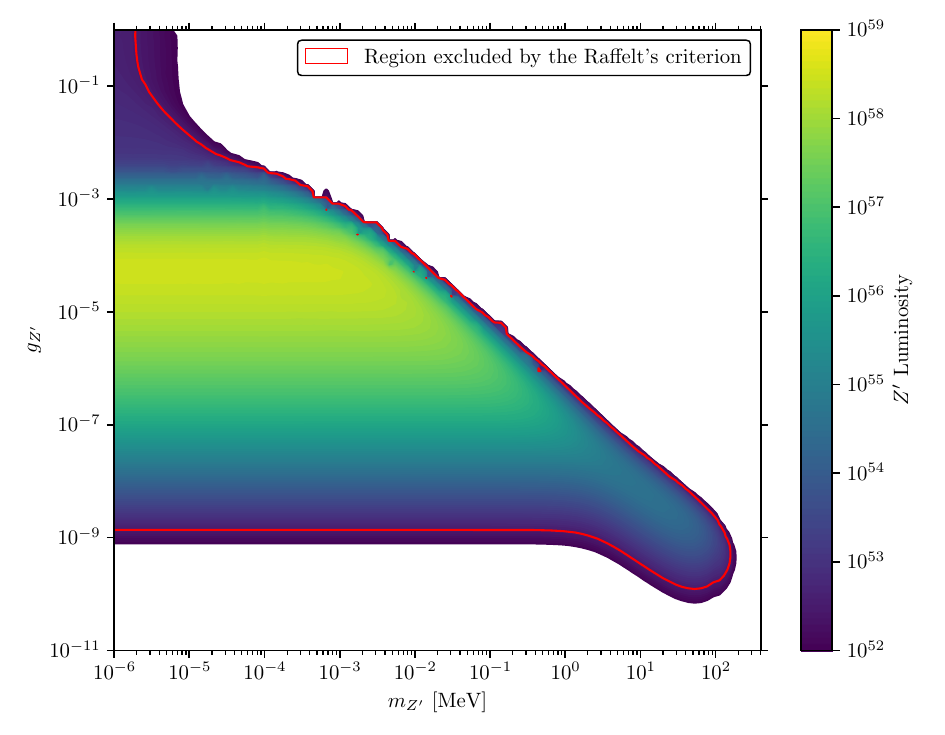}%
		\includegraphics[width=.5\columnwidth]{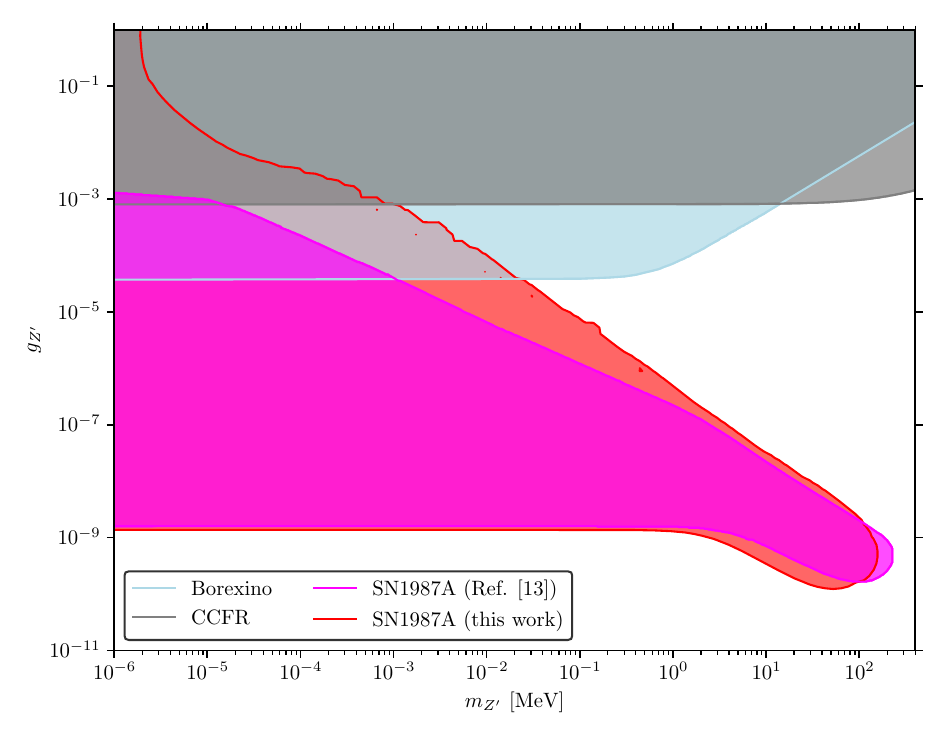}%
		\caption{(Left) The luminosity contour region for $L_{Z'}>10^{52}$ [erg/s]. The region corresponds to $L_{Z'}>L_{\text{Raffelt}}=3\times 10^{52}$ [erg/s] is enclosed by the orange line. (Right) The excluded parameter region due to existing experiments and the previous work~\cite{Croon:2020lrf}. Our result is shaded in orange colour. %Read the context below for a more detail explanation.
  }%
		\label{fig:Contour}%
	\end{figure}

%The details for calculating this limiting luminosity, denoted by $L_{\infty}$, are given in Fig.~\ref{fig.Sphere_e} and Eq.~\eqref{eq.Linf}. The expected plateau value $L_{\infty}$ as a function of $m_{Z'}$ is given in the right panel of Fig.~\ref{fig:Luminosity}. 
With the physics of $Z^{\prime}$ luminosity in the large coupling regime clarified, we once more turn our attention to Fig.~\ref{fig:Luminosity}. On the left panel of the figure, a full numerical calculation results in a plateau value of approximately $3\times10^{52}$ [erg/s] at $m_{Z'}=2$ eV, and this plateau value is accurately predicted by the right panel of the figure. Furthermore, when the mass of $Z'$ falls below approximately $2$ eV, $L_{\infty}$ exceeds the upper bound set by the Raffelt criterion. This implies that for $m_{Z'}<2$ eV, the excluded parameter region will no longer be bounded from above by the trapping limit. Thus, the plateau phenomenon significantly constrains the model when $m_{Z'}<2$ eV. %This striking result is 
This interesting result is demonstrated in the complete contour plot shown in Fig.~\ref{fig:Contour}. Before discussing the significance of Fig.~\ref{fig:Contour}, there is a caveat worth mentioning as $\alpha_{Z'}$ approaches or even surpasses $\alpha$. In such a case, the interaction rate for $Z^{\prime}\mu^-\to Z^{\prime}\mu^-$ is comparable or even faster than that of $Z^{\prime}\mu^-\to \gamma \mu^-$. Hence the former process should be considered for the propagation of $Z^{\prime}$ in addition to those reabsorption processes considered in Fig.~\ref{fig.reabsorptionprocesses}. Consequently the $Z^{\prime}$ trajectory inside the neutrinosphere is no longer a straight line. We shall not consider such a scenario in this work. Thus one should disregard constrained regions with $g_{Z'}>0.1$.
	
On the left panel of Fig.~\ref{fig:Contour}, we present the luminosity contour plot by scanning through the parameter space $(m_{Z'},g_{Z'})$ in Eq.~\eqref{eq.LuminosityZ'}. The orange line encloses the excluded parameter region where $Z'$ luminosity $L_{Z'} > L_{\text{Raffelt}}$. As $m_{Z'}$ approaches $\sim 2$ eV from above, the upper orange curve tends towards a vertical line, which excludes all parameter space $g_Z'$.  %The plateau phenomenon causes the trapping limit to disappear. 
Precisely speaking, for $m_{Z'}<2$ eV, the plateau value in Fig.~\ref{fig:Luminosity} is already greater than the Raffelt bound, resulting in the vanishing of the trapping limit.
Our work (shaded in orange) extends the trapping limit gradually as $m_{Z'}$ decreases to $\sim 2$ eV. In comparison to the result of Croon {\it et al.}~(the magenta shaded region in the right panel of Fig.~\ref{fig:Contour}), our result represents a significant extension of excluded parameter region. This difference is largely due to the extra attenuation region between the radii $R_{\nu}$ and $R_{\rm far}$ adopted in~\cite{Croon:2020lrf}. On the other hand, even in the $R_{\rm far}=R_{\nu}$ case, exclusion regions resulting from two calculations might still be different as hinted by the difference in $L_{Z'}$ for $m_{Z'}=0.1$ eV mentioned in the beginning of the section.     

We have so far neglected plasma corrections to the $Z'$ production and attenuation processes. %It is important to note that such corrections must be considered when the plasma frequency $\omega_p$ is comparable or greater than $m_{Z^{\prime}}$. %Here, we discuss the correction on loop-Bremsstrahlung and other processes that couple to neutrinos separately. 
  We shall focus on effects of such corrections to the trapping regime constraints. Since the determination of
  trapping regime constraints only involves reactions occurring near the surface of neutrinosphere, we simply consider $Z^{\prime}$ decay, inverse loop bremsstrahlung and their inverse processes while neglect semi-Compton due to the vanishing muon number density. 

  For $Z^{\prime}\to \nu\bar{\nu}$ or its inverse process, $Z^{\prime}$ acquires a plasma mass $\omega^{Z'}_p = \sqrt{4\pi \alpha_{Z'} n_\nu/E_\nu} \approx g_{Z'}T_{\nu} \sqrt{2\zeta(3)/(2\pi^2)} \approx 0.35g_{Z'}T_{\nu}$~\cite{Croon:2020lrf} from its self-energy contributed by ambient neutrinos. Since $T_\nu\sim 3$ MeV near the neutrinosphere, we have $(\omega^{Z'}_p/{\rm MeV})\approx g_{Z'}$. It is important to note that the $Z'$ mass should be corrected when the plasma frequency $\omega_p^{Z'}$ is comparable or greater than $m_{Z^{\prime}}$. Those relevant  $\omega_p^{Z'}$ values (depending on $g_{Z'}$) for such comparisons can be 
obtained by taking $g_Z'$ values from the upper orange curve on the left panel of Fig.~\ref{fig:Contour}. This choice is self-consistent because one needs to determine if the upper constraint curve should be shifted due to the plasma correction. One can see that $\omega^{Z'}_p \geq m_{Z'}$ for $m_{Z'}\leq 1$ keV. Hence for $m_{Z'}\leq 1$ keV, the plasma corrections to $Z^{\prime}\to \nu\bar{\nu}$ and its inverse process should be considered. In this case, the effective mass of $Z^{\prime}$ becomes $m_{Z'}^{\rm eff}\equiv \sqrt{m_{Z'}^2+(\omega^{Z'}_p)^2}$ for both transverse and longitudinal modes in the limit $\vert \mathbf{k}\vert \to 0$ where $\mathbf{k}$
  is the three-momentum of $Z'$.  The $Z'$ self-energy also introduces wave function renormalizations or the so-called residual functions, which are denoted as $Z_T$ and $Z_L$ for  transverse and longitudinal modes, respectively. Using the approaches in~\cite{Weldon:1982aq,Braaten:1993jw}, we obtain $Z_{T,L}=1+\mathcal{O}(\vert \mathbf{k}\vert ^2(\omega^{Z'}_p)^2/\omega^4 )$. The corrections $\delta Z_{T,L}\equiv Z_{T,L}-1$ are negligible since $\omega\simeq \omega^{Z'}_p$ and $\vert \mathbf{k} \vert \ll \omega$ in the current limit.  In the limit $\omega, \vert \mathbf{k} \vert \gg \omega^2-\vert \mathbf{k} \vert ^2$ with $\vert \mathbf{k} \vert/\omega\to 1$, both the amplitude of  $Z^{\prime}\to \nu\bar{\nu}$ and that of $\nu\bar{\nu}\to Z'$ for longitudinal $Z'$ are vanishing. To see this, we note that $\epsilon^{\mu}_{Z', L}\equiv ( \mathbf{k}, \omega \hat{\mathbf{k}} )/\sqrt{\omega^2-\vert \mathbf{k}\vert^2}=k^{\mu}/\sqrt{\omega^2-\vert \mathbf{k}\vert^2}
  +\mathcal{O}(\sqrt{\omega^2-\vert \mathbf{k}\vert^2}/\omega)$. This implies $\epsilon^{\mu}_{Z', L}\bar{u}(p_1)\gamma_{\mu}P_L v(p_2)\simeq k^{\mu}\bar{u}(p_1)\gamma_{\mu}P_L v(p_2)/\sqrt{\omega^2-\vert \mathbf{k}\vert^2} \to 0$ (for $Z'\to \nu\bar{\nu}$) 
  with $p_1$ and $p_2$ the momenta of outgoing $\nu$ and $\bar{\nu}$. For the transverse mode, it can be shown that $m_{Z'}^{\rm eff}=\sqrt{m_{Z'}^2+(m^{Z'}_T)^2}$ with $\omega^{Z'}_p\leq m^{Z'}_T\leq \sqrt{3/2}\omega^{Z'}_p$ depending on the velocity of ambient 
  electrons~\cite{Braaten:1993jw}. We can also show that the correction to the residual function in this limit is $\delta Z_T=\mathcal{O}((\omega^{Z'}_p)^2/k^2)$, which is also negligible.  
  Clearly, for $m_{Z'}\leq 1$ keV, the plasma corrections to $Z'\to \nu\bar{\nu}$ and its inverse process amount to replacing $m_{Z'}$ by $m^{\rm eff}_{Z'}$
  which is either $\sqrt{m_{Z'}^2+(\omega^{Z'}_p)^2}$ or $\sqrt{m_{Z'}^2+(m^{Z'}_T)^2}$ depending on the kinematics. 
  From Eqs.~(\ref{eq.pairCoalnu}) and (\ref{eq.labframedecayrate_neutrino}), the interaction rates of both $\nu\bar{\nu}\to Z'$ and $Z'\to \nu\bar{\nu}$ are enhanced by the same factor 
  $(m_{Z'}^{\rm eff})^2/m_{Z'}^2$. %The combined effect to the luminosity $L_{Z'}$ can be seen from Eqs.~(\ref{eq.Abar}) and (\ref{eq.LuminosityZ'}) where $\Gamma_{\rm abs}\to (m_{Z'}^{\rm eff})^2/m_{Z'}^2\cdot \Gamma_{\rm abs}$ and $\Gamma_{\rm prod}\to (m_{Z'}^{\rm prod})^2/m_{Z'}^2\cdot \Gamma_{\rm prod}$.
  %This generally leads to a suppression of $L_{Z'}$ since the resulting exponential suppression of the former dominates the enhancement of the latter. Consequently, the trapping regime constraints become less stringent due to thermal corrections. 
  %We separate the thermal corrections into two categories. 
  %The first category includes corrections to pair coalescence and its inverse process. 
  %The first category includes corrections to the loop-Bremsstrahlung and its inverse process while the second category contain corrections to pair coalescence, semi-Compton and their inverse processes. 
  
%For $m_{Z'}< 0.02$ keV, the inverse loop-bremsstrahlung process dominates $Z^{\prime}\to \nu\bar{\nu}$ (same is true for the inverse process due to the principle of detailed balance). 
For the inverse loop bremsstrahlung process, the major plasma correction appears in the photon propagator that mixes with external $Z^{\prime}$ line. The correction of this type has been discussed in An {\it et al.}~\cite{An:2013yfc}, which gains an overall extra factor $(m_{Z'}/\omega^{\gamma}_p)^n$ for $m_{Z'}< \omega^{\gamma}_p$, with $n=2$ for longitudinal modes and $n=4$ for transverse modes. The photon plasma frequency is given by $\omega^{\gamma}_p = \sqrt{4\pi \alpha_{\rm EM} n_e/E_e}$ with $E_e^2=m_e^2+(3\pi^2 n_e)^{2/3}$. Numerically, we have $\omega^{\gamma}_p\approx 0.3$ MeV on the surface of neutrinosphere according to the simulation in~\cite{Bollig:2020xdr}. %Contrast to the $\nu\bar{\nu}\to Z'$/$Z^{\prime}\to \nu\bar{\nu}$ case, the suppression of both $n+p\to n+p+Z'$ and its inverse process lead to the enhancement of $L_{Z'}$. As a result, the trapping regime constraints for $m_{Z'}< 0.02$ keV become even more stringent.  

The overall plasma corrections to $L_{Z'}$ begin from $m_{Z'}\leq \omega^{\gamma}_p$ with $\omega^{\gamma}_p=0.30$ MeV on the surface of neutrinosphere as just stated.
For $4.6 \, {\rm keV}\leq m_{Z'}\leq 0.30 \, {\rm MeV}$, $Z'\rightarrow \nu + \bar{\nu}$ dominates inverse loop bremsstrahlung before including the plasma corrections. Along the upper orange curve on the left panel of Fig.~\ref{fig:Contour}, one can see that $g_{Z'} < 10^{-3}$ in this mass range, which implies $\omega^{Z'}_p < 1$ keV. 
Hence only $Z'+n+p\to n+p$ and its inverse process are suppressed by plasma corrections while  $Z^{\prime}\to \nu\bar{\nu}$ and $\nu\bar{\nu}\to Z'$ are not affected. Since the former processes are subleading, the overall plasma correction to $L_{Z'}$ is not significant. For $1.0 \, {\rm keV}\leq m_{Z'}\leq 4.6 \, {\rm keV}$, the relevant $\omega^{Z'}_p$ remains no larger than $1$ keV. Hence, $Z^{\prime}\to \nu\bar{\nu}$ and its inverse process remain unaffected while $Z'+n+p\to n+p$ and its inverse process are suppressed by plasma corrections. Since both the production and attenuation rates of $Z'$ are suppressed, their combined effect to $L_{Z'}$ requires a carful study. For $m_{Z'}\leq 1$ keV, both type of processes receive plasma corrections but in opposite directions, i.e., the event rate of $Z' \to \nu\bar{\nu} $ and that of its inverse process are enhanced while event rates of both $Z'+n+p\to n+p$ and $n+p\to n+p+Z'$ are suppressed. The overall 
 effects to $L_{Z'}$ once more requires a careful study. We shall tackle these issues in future publications.
 %For other processes that couple directly to neutrinos (tree-level), the correction is applied to the coupling between $Z'$ and $\nu_{\mu/\tau}$. In this case, the plasma frequency is $\omega_p = \sqrt{4\pi \alpha_{Z'} n_\nu/E_\nu} \approx g_{Z'}T_{\nu} \sqrt{2\zeta(3)/(2\pi^2)} \approx 0.35g_{Z'}T_{\nu}$. Roughly speaking, the thermal correction cannot be neglected when $m_{Z'} \ll g_{Z'}T_{\nu}$. 
% For more details, see references \cite{An:2013yfc,Croon:2020lrf}.
 	
%Furthermore, for $m_{Z'}<2$ eV, the plateau value in Fig.~\ref{fig:Luminosity} is already greater than the Raffelt bound, resulting in the vanishing of the trapping limit. 

Our analysis shows that the extended exclusion region of the parameter space overlaps with the excluded region derived from neutrino trident experiment in CCFR~\cite{Altmannshofer:2014pba} at a $95\%$ confidence level, as well as 
%the muon anomalous magnetic moment measurement in 
the exclusion region derived from the
Borexino solar neutrino measurement data~\cite{Amaral:2020tga}. Therefore, our findings confirm that this overlapped parameter region is indeed 
disfavored.
%not allowed.
	%Besides that, we get pretty much agreements with Croon {\it et al.}. Supernova neutrinos give new constraints on the $U(1)_{L_\mu-L_\tau}$ model and have a huge potential in providing constraints for other new long-living weak interacting hypothetical particles.

\section{summary and conclusions}\label{sec.05}

We pointed out that SN constraints on BSM theories involving extra neutral bosons near the trapping regime should be handled with care. Specifically, we suggest that the attenuation boundary for the new boson should be taken to be identical to the production boundary, and the luminosity of the new boson will tend to a constant value instead of decreasing to zero at a large coupling limit. We addressed this as the plateau phenomenon. It has been argued in~\cite{Croon:2020lrf} that $R_{\rm far}$ should be considered a value greater than $R_{\nu}$. However, we asserted that this is not self-consistent as the extra neutral boson can still be produced within the region between $R_\nu$ and $R_{\rm far}$. %One cannot stress which of these processes is more important. 

To demonstrate the aforementioned plateau phenomenon, we took the $U(1)_{L_\mu-L_\tau}$ model as an illustrative example.
We employed the luminosity formula Eq.~(\ref{eq.ULum}) from Chang {\it et al.}~\cite{Chang:2016ntp} to unify the calculations of free-streaming and trapping regimes. %Unlike previous studies~\cite{Chang, Croon:2020lrf}, 
However, we did not use the approximation employed by~Ref.\cite{Chang:2016ntp} for computing the attenuation factor (see comments below Eq.~(\ref{eq.LuminosityZ'})), and we computed it numerically with $R_{\rm far}=R_{\nu}$. By performing complete numerical calculations and theoretical derivations, we verified and derived the plateau phenomenon. Notably, for the $U(1)_{L_\mu - L_\tau}$ model, we found that $L_\infty (m_{Z'})$ exceeded %the SN neutrino allowed value when 
$L_{\rm Raffelt}$ for $m_{Z'}<2$ eV. %which differs from the results reported by Croon {\it et al.}~\cite{Croon:2020lrf}. 
Although Ref.~\cite{Croon:2020lrf} also presented the $Z'$ boson luminosity at the limit $R_{\rm far}\to R_{\nu}$ for $m_{Z'}=0.1$ eV, their result does not  show the plateau phenomenon at the large $g_{Z'}$ limit. 
We have discussed plasma corrections to the constraint curve in the trapping regime. We have shown that such corrections might be non-negligible for $m_{Z'}< 4.6$ keV. We leave the detailed study of this issue to future publications. Before closing, we like to point out that our work does not consider the extra BSM neutral boson as the portal between visible and dark sectors. For such a scenario and the corresponding SN1987A bounds to the model, see~\cite{Sung:2021swd,Manzari:2023gkt}.     

In conclusion, we have stressed the significance of giving equal consideration to the region of absorption processes as that of production processes. At the boundary of the neutrinosphere, the interplay between emissivity and attenuation factor results in a constant luminosity, denoted as $L_\infty (m)$, instead of its gradual decrease to zero in the event of a large coupling. Our analysis hence extends the exclusion region of $U(1)_{L_\mu-L_\tau}$ parameter space in comparison to the analysis by~\cite{Croon:2020lrf}.
%Depending on the mass of the BSM particle, $L_\infty (m)$ could potentially exceed the %experimental 
%allowable luminosity value from SN observations and thus, strongly restrict the parameter space of BSM models. 

\section*{Acknowledgements}
We kindly thank E.~Vitagliano for pointing out 
%valuable insights on the 
recent updates concerning modified luminosity calculations~\cite{Caputo:2021rux,Caputo:2022rca,Bollig:2020xdr} and K.~Akita for highlighting an %interesting 
alternative approach in constraining similar models~\cite{Akita:2023iwq, Fiorillo:2022cdq, Jegerlehner:1996kx, Mirizzi:2005tg}.
We also thank M.-R. Wu for discussions on strong $\alpha_{Z'}$ limit as mentioned in section \ref{sec.04}. This work is supported by National Science and Technology Council, Taiwan, under Grant Nos. NSTC 111-2112-M-A49-027 and NSTC 112-2112-M-A49-017.

\newpage
\appendix

\section{\texorpdfstring{$Z'$}{} decay and lepton pair coalescence}\label{sec.Decay_pC}
This section will discuss four calculation processes in detail, comprising two $Z'$ decay channels and two $Z'$ production channels involving muon and neutrino pair coalescence, as shown by Fig.~\ref{fig.pC_mu_nu}. We will commence by calculating the decay rates for $Z'\rightarrow \mu^+ + \mu^-$ and $Z'\rightarrow \nu + \bar{\nu}$.

\noindent
\begin{minipage}{1.0\columnwidth}%
	\vspace{-0.8em}
	\centering
\begin{minipage}{.3333\columnwidth} %don't touch this unless you really know what you're doing
	\begin{figure}[H]
		\centering
		\begin{fmffile}{decaymu}
			\setlength{\unitlength}{.8\columnwidth}
			\vspace{1em}
			\begin{fmfgraph*}(0.9,0.5)
				%\fmfstraight
				\fmfpen{1.0}
				\fmfleft{i1}
				\fmfright{o1,o2}
				\fmf{fermion}{o2,v1}
				\fmf{fermion}{v1,o1}
				\fmf{photon,tension=1.5}{i1,v1}
				\fmfv{l=$\mu^-$, l.a=0}{o2}
				\fmfv{l=$\mu^+$, l.a=0}{o1}
				\fmfv{l=$Z'$, l.a=180}{i1} 
			\end{fmfgraph*}%
		\end{fmffile}%
		\caption*{$Z'\rightarrow \mu^+ + \mu^-$}
	\end{figure}%
\end{minipage}%
\begin{minipage}{.25\columnwidth}%
	\begin{equation*}
		\overset{m_{\mu}\rightarrow 0}{\underset{\Gamma_{Z'\rightarrow \mu^- \mu^+} \text{ to } \Gamma_{Z'\rightarrow \nu_{\mu} \bar{\nu}_{\mu}}}{\xrightarrow{{\makebox[.5\columnwidth]{}}}}}%
	\end{equation*}
\end{minipage}%
\begin{minipage}{.3333\columnwidth} %don't touch this unless you really know what you're doing
	\begin{figure}[H]
		\centering
		\begin{fmffile}{decaynu}
			\setlength{\unitlength}{.8\columnwidth}
			\vspace{1em}
			\begin{fmfgraph*}(0.9,0.5)
				%\fmfstraight
				\fmfpen{1.0}
				\fmfleft{i1}
				\fmfright{o1,o2}
				\fmf{fermion}{o2,v1}
				\fmf{fermion}{v1,o1}
				\fmf{photon,tension=1.5}{i1,v1}
				\fmfv{l=$\nu$, l.a=0}{o2}
				\fmfv{l=$\bar{\nu}$, l.a=0}{o1}
				\fmfv{l=$Z'$, l.a=180}{i1} 
			\end{fmfgraph*}%
		\end{fmffile}%
		\caption*{$Z'\rightarrow\nu+\bar{\nu}$}
	\end{figure}%
\end{minipage}%
\end{minipage}%
\vspace{1em}

The square of the amplitude for the non-polarized decay of $Z'$ into a muon pair, $Z' \rightarrow \mu^+ + \mu^-$, is provided by
\begin{align}
	\langle \vert \mathcal{M} \vert^2 \rangle &\equiv \frac{1}{3}\sum_{\text{pol.}} \vert \mathcal{M} \vert^2 = \frac{4}{3}g_{Z'}^2 (2m_\mu^2+m_{Z'}^2). \label{aeq.Msquared_decayZp_muon}
\end{align}
Here, the pre-factor $1/3$ incorporates the spin average of $Z'$. The calculation of the decay rate, performed in the rest frame of the initial particle $Z'$, follows this expression:
\begin{align}
	\Gamma_{Z'\rightarrow \mu^- + \mu^+} &= \frac{p^*}{32\pi^2 m_{Z'}^2} \int \langle \vert \mathcal{M} \vert^2 \rangle d\Omega \label{aeq.restframedecayrate01}\\
	&= \frac{g_{Z'}^2 m_{Z'}}{12\pi} \left( 1+\frac{2m_{\mu}^2}{m_{Z'}^2} \right) \sqrt{1-\frac{4m_{\mu}^2}{m_{Z'}^2}}, \label{aeq.restframedecayrate}
\end{align}
in which $p^* = \sqrt{m_{Z'}^2 - 4m_\mu^2}/2$. To derive the lab frame decay rate in thermal equilibrium, multiply Eq.~\eqref{aeq.restframedecayrate} by an inverse special relativity gamma factor, denoted as $1/\gamma = m_{Z'}/\omega$. Additionally, consider the \emph{Pauli-blocking factor}, denoted as $1-f_{\text{F}}(\omega/2,\mu)$ (where $\mu$ denotes the chemical potential of the corresponding particle), which accounts for the presence of final state muons. The decay rate in the laboratory frame can be expressed as
\begin{align}
	\Gamma_{Z'\rightarrow \mu^- + \mu^+}^{\text{lab}}
	&= \frac{g_{Z'}^2 m_{Z'}^2}{12\pi\omega} \left( 1+\frac{2m_{\mu}^2}{m_{Z'}^2} \right) \sqrt{1-\frac{4m_{\mu}^2}{m_{Z'}^2}}\times (1-f_{\text{F}}(\omega/2,\mu_{\mu^-}))(1-f_{\text{F}}(\omega/2,\mu_{\mu^+})),
\end{align}
where $\omega$ represents the energy of the newly introduced gauge boson $Z'$, and $\mu_{\mu^-}$/$\mu_{\mu^+}$ denotes the chemical potential of $\mu^-$/$\mu^+$. Since the chemical potential is contingent on the simulation, the term related to the Pauli-blocking factor, denoted as $(1-f_{\text{F}}(\omega/2,\mu_{\mu^-}))(1-f_{\text{F}}(\omega/2,\mu_{\mu^+}))$, should not be disregarded in principle. For $\mu^+$, the temperature of SNe is typically a few tenths of a MeV, significantly smaller than the muon mass, i.e., $T\ll m_\mu<\omega$. Consequently,
\begin{align*}
	1-f_{\text{F}}(\omega/2,\mu_{\mu^+}) &\equiv  1-\frac{1}{e^{(\omega-\mu_{\mu^+})/T}+1} \\
	&= 1-\frac{1}{e^{(\omega+\mu_{\mu^-})/T}+1} \\
	&\approx 1.
\end{align*}
The final approximation remains valid because the values of $\omega$ and $\mu_{\mu^-}$ in the numerator of the exponential function surpass the hundred MeV scale, making them significantly larger than the temperature $T$, which is only a few tenths of the MeV scale. In other words, $\exp[(\omega+\mu_{\mu^-})/T]\gg 1$. Conversely, the dependence of $(1-f_{\text{F}}(\omega/2,\mu_{\mu^-}))$ is on the simulation data of the muon chemical potential $\mu_{\mu^-}$. Despite being reliant on simulation data, the magnitude of this quantity is constrained:
\begin{align}
	1 &\geq 1-f_{\text{F}}(\omega/2,\mu_{\mu^-}) \geq \frac{1}{2}.
\end{align}
Since a factor between $1$ and $1/2$ will not affect the overall order of magnitude, in this work $(1-f_{\text{F}}(\omega/2,\mu_{\mu^-})) \approx 1$ has been applied for simplicity in calculations. Ultimately, the decay rate of $Z'\rightarrow \mu^- + \mu^+$ in the lab frame is determined as
\begin{equation}
	\Gamma_{Z'\rightarrow \mu^- + \mu^+}^{\text{lab}}
	\approx \frac{g_{Z'}^2 m_{Z'}^2}{12\pi\omega} \left( 1+\frac{2m_{\mu}^2}{m_{Z'}^2} \right) \sqrt{1-\frac{4m_{\mu}^2}{m_{Z'}^2}} 
	\label{aeq.labframedecayrate_muon}
\end{equation}

For $Z'\rightarrow \nu + \bar{\nu}$, only left-handed neutrinos and right-handed anti-neutrinos can play a role. Consequently, the squared matrix element of $Z'\rightarrow \nu_{\mu} + \bar{\nu}_\mu$ is half that of $Z'\rightarrow \mu^- + \mu^+$. Taking $m_\mu \rightarrow 0$, we obtain the expression
\begin{align}
	\langle \vert \mathcal{M} \vert^2 \rangle &\equiv \frac{1}{3}\sum_{\text{pol.}} \vert \mathcal{M} \vert^2 = \frac{2}{3}g_{Z'}^2 m_{Z'}^2.
    \label{aeq.Msquared_decayZp_neutrino}
\end{align}
The decay rate of $Z'\rightarrow \nu_\mu + \bar{\nu}_\mu$ in the lab frame can be obtained by inserting Eq.~\eqref{aeq.Msquared_decayZp_neutrino} into Eq.~\eqref{aeq.restframedecayrate01}. Result in
\begin{align}
    \Gamma_{Z'\rightarrow \nu_\mu + \bar{\nu}_\mu}^{\text{lab}} &= \frac{g_{Z'}^2 m_{Z'}^2}{24\pi\omega}.
\end{align}
Hence, taking into account the production of $Z'$ from both $\nu_\mu$ and $\nu_\tau$, the overall production rate amounts to
\begin{equation}
		\Gamma_{Z'\rightarrow \nu + \bar{\nu}}^{\text{lab}}
		\approx \frac{g_{Z'}^2 m_{Z'}^2}{12\pi\omega}.
	\label{aeq.labframedecayrate_neutrino}
\end{equation}
It's important to note that neutrinos differ significantly from muons, not only in terms of mass but also due to the impact of helicity states on the overall prefactor. Taking $m_\mu\rightarrow 0$ in Eq.~\eqref{aeq.Msquared_decayZp_muon} leads to Eq.~\eqref{aeq.labframedecayrate_neutrino} is merely a coincidence.

\noindent
\begin{minipage}{1.0\columnwidth}%
	\begin{figure}[H]%
		\centering
		\vspace{-1.0em}
		\begin{minipage}{.43333\columnwidth} %don't touch this unless you really know what you're doing
			\begin{figure}[H]
				\centering
				\begin{fmffile}{paircoalescence_mu}
					\setlength{\unitlength}{.8\columnwidth}
					\vspace{1em}
					\begin{fmfgraph*}(0.9,0.5)
						%\fmfstraight
						\fmfpen{1.0}
						\fmfleft{i1,i2}
						\fmfright{o1}
						\fmf{fermion}{i2,v1}
						\fmf{fermion}{v1,i1}
						\fmf{photon,tension=1.5}{v1,o1}
						\fmfv{l=$\mu^-$, l.a=180}{i2}
						\fmfv{l=$\mu^+$, l.a=180}{i1}
						\fmfv{l=$Z'$, l.a=0}{o1} 
					\end{fmfgraph*}%
				\end{fmffile}%
				\caption*{$\mu^- + \mu^+ \rightarrow Z'$}
			\end{figure}%
		\end{minipage}%
		\qquad\quad
		\begin{minipage}{.43333\columnwidth}
			\begin{figure}[H]
				\centering
				\begin{fmffile}{paircoalescence_nu}
					\setlength{\unitlength}{.8\columnwidth}
					\vspace{1em}
					\begin{fmfgraph*}(0.9,0.5)
						%\fmfstraight
						\fmfpen{1.0}
						\fmfleft{i1,i2}
						\fmfright{o1}
						\fmf{fermion}{i2,v1}
						\fmf{fermion}{v1,i1}
						\fmf{photon,tension=1.5}{v1,o1}
						\fmfv{l=$\nu_\mu,,\nu_\tau$, l.a=180}{i2}
						\fmfv{l=$\bar{\nu}_\mu,,\bar{\nu}_\tau$, l.a=180}{i1}
						\fmfv{l=$Z'$, l.a=0}{o1} 
					\end{fmfgraph*}%
				\end{fmffile}%
				\caption*{$\nu + \bar{\nu} \rightarrow Z'$}
			\end{figure}%
		\end{minipage}%
		%\vspace{1.0em}
		\caption{Feynman diagrams of muon and neutrino pair coalescence.}%
		\label{fig.pC_mu_nu}
	\end{figure}%
\end{minipage}\vspace{12pt plus 2pt minus 2pt}%

According to the principle of detailed balance, the pair-coalescence rate $\Gamma_{\mu^- + \mu^+ \rightarrow Z'}$ is simply related to $\Gamma_{Z'\rightarrow \mu^- + \mu^+}^{\text{lab}}$ (Eq.~\eqref{aeq.labframedecayrate_muon}) by
\begin{align}
	\Gamma_{\mu^- + \mu^+ \rightarrow Z'} &= \frac{3}{4} \times \Gamma_{Z'\rightarrow \mu^- + \mu^+}^{\text{lab}}\times 4 \times e^{-\omega/T}\nonumber\\
	&\approx \frac{g_{Z'}^2 m_{Z'}^2}{4\pi\omega} \left( 1+\frac{2m_{\mu}^2}{m_{Z'}^2} \right) \sqrt{1-\frac{4m_{\mu}^2}{m_{Z'}^2}} e^{-\omega/T},
\end{align}
in which the pre-factor $3/4$ and $4$ account for the correct polarization averaging in the matrix element squared and the degeneracy factor in reversing the process by the principle of detailed balance. 
\begin{comment}
    Finally, the production of the final state particle $Z'$ is further affected by the \emph{Bose-enhancement factor} $1+f_{\text{B}}(\omega)$ in thermal equilibrium environment
\begin{align}
	\text{Pauli-blocking factor : }& 1-f_{\text{F}}(\omega) = 1-\frac{1}{e^{\omega/T}+1} \\
	\text{Bose-enhancement factor : }& 1+f_{\text{B}}(\omega) = 1+\frac{1}{e^{\omega/T}-1}
\end{align}
This results in the final production rate
\begin{equation}
	\Gamma_{\mu^- + \mu^+ \rightarrow Z'} \approx \frac{g_{Z'}^2 m_{Z'}^2}{4\pi\omega} \left( 1+\frac{2m_{\mu}^2}{m_{Z'}^2} \right) \sqrt{1-\frac{4m_{\mu}^2}{m_{Z'}^2}} \times e^{-\omega/T} \times \left( 1 + \frac{1}{e^{\omega/T}-1} \right)
\end{equation}
\end{comment}
Note that the Pauli-blocking factor is ignored for the same reason as Eq.~\eqref{aeq.labframedecayrate_muon}. It gives the overall formula
\begin{equation}
	\Gamma_{\mu^- + \mu^+ \rightarrow Z'} \approx \frac{g_{Z'}^2 m_{Z'}^2}{4\pi\omega} \left( 1+\frac{2m_{\mu}^2}{m_{Z'}^2} \right) \sqrt{1-\frac{4m_{\mu}^2}{m_{Z'}^2}} \times e^{-\omega/T}  .\label{aeq.pairCoalmu}
\end{equation}

According to the principle of detailed balance, the pair-coalescence rate $\Gamma_{\nu_\mu + \bar{\nu}_\mu \rightarrow Z'}$ is simply related to $\Gamma_{Z'\rightarrow \nu_{\mu} + \bar{\nu}_{\mu}}^{\text{lab}}$ by
\begin{align}
	\Gamma_{\nu_\mu + \bar{\nu}_\mu \rightarrow Z'} &= 3 \Gamma_{Z'\rightarrow \nu_{\mu} + \bar{\nu}_{\mu}}^{\text{lab}} e^{-\omega/T} \nonumber\\
	&\approx \frac{g_{Z'}^2 m_{Z'}^2}{8\pi \omega} e^{-\omega/T}.
\end{align}
Since the neutrino is always left-handed, the number of initial neutrino polarization combinations is $1$, and we do not divide the first equality by a factor of $4$. Taking into consideration the production from $\nu_\tau + \bar{\nu}_\tau \rightarrow Z'$, the production rate of the new gauge boson $Z'$ through the pair-coalescence of $\nu + \bar{\nu} \rightarrow Z'$ is
\begin{equation}
	\Gamma_{\nu + \bar{\nu} \rightarrow Z'} \approx \frac{g_{Z'}^2 m_{Z'}^2}{4\pi \omega} \times e^{-\omega/T}. \label{aeq.pairCoalnu}
\end{equation}

\section{\texorpdfstring{$Z'$}{} production and absorption in semi-Comption processes}\label{sec.SemiComp}
We obtain a simple production rate by investigating and modifying the well-known Compton process. Since both diagrams are very similar, the semi-Compton cross-section of $Z'$ production could be estimated by the usual Compton process. 

\noindent
\begin{minipage}{.4\columnwidth} 
	\begin{figure}[H]
		\centering
		\begin{fmffile}{compton}
			\setlength{\unitlength}{.6666\columnwidth}
			\vspace{1em}
			\begin{fmfgraph*}(0.9,0.5)
				\fmfpen{1.0}
				\fmfleft{i1,i2}
				\fmfright{o1,o2}
				\fmf{fermion}{i2,v2}
				\fmf{fermion}{v1,o1}
				\fmf{photon}{v2,o2}
				\fmf{photon}{i1,v1}
				\fmfv{l=$\mu^-$, l.a=180}{i2}
				\fmfv{l=$\gamma$, l.a=180}{i1}
				\fmfv{l=$\gamma$, l.a=0}{o2}
				\fmfv{l=$\mu^-$, l.a=0}{o1} 
				\fmf{fermion,label=,tension=0}{v2,v1} 
			\end{fmfgraph*}%
		\end{fmffile}%
		\caption*{Compton}
	\end{figure}%
\end{minipage}%
\begin{minipage}{0.2\columnwidth} %requires mpost command
\begin{equation*}
	\overset{\text{Replace } \gamma \text{ with } Z'}{\xrightarrow{{\makebox[.9\columnwidth]{}}}}%
\end{equation*}%
\end{minipage}%
\begin{minipage}{0.4\columnwidth} %requires mpost command
	\begin{figure}[H]
		\centering
		\begin{fmffile}{semicompton}
			\setlength{\unitlength}{.6666\columnwidth}
			\vspace{1em}
			\begin{fmfgraph*}(0.9,0.5)
				\fmfpen{1.0}
				\fmfleft{i1,i2}
				\fmfright{o1,o2}
				\fmf{fermion}{i2,v2}
				\fmf{fermion}{v1,o1}
				\fmf{photon}{v2,o2}
				\fmf{photon}{i1,v1}
				\fmfv{l=$\mu^-$, l.a=180}{i2}
				\fmfv{l=$\gamma$, l.a=180}{i1}
				\fmfv{l=$Z'$, l.a=0}{o2}
				\fmfv{l=$\mu^-$, l.a=0}{o1} 
				\fmf{fermion,label=,tension=0}{v2,v1} 
			\end{fmfgraph*}%
		\end{fmffile}%
		\caption*{Semi-Compton}
	\end{figure}%
\end{minipage}%
\vspace{1em}

By the Klein-Nishina formula, the scattering cross-section is given by
\begin{align}
	\frac{d\sigma_{\text{C}}}{d\cos\theta} &= \frac{\pi\alpha^2}{m_\mu^2}\left(\frac{\omega'}{\omega}\right)\left(\frac{\omega'}{\omega}+\frac{\omega}{\omega'}-\sin^2\theta \right),
\end{align}
where 
\begin{equation}
    \omega' = \frac{\omega}{1+\frac{\omega}{m_\mu}(1-\cos\theta)},
\end{equation}
in which $\omega$ and $\omega'$ represent the energy of initial and final state photons. They are connected by the energy and momentum conservation law. assuming the photon energy is much less than ff muon, $\omega\ll m_\mu$, the cross-section can be written in a simpler form
\begin{align}\label{eq.sCsigma}
	\frac{d\sigma_{\text{C}}}{d\cos\theta} &\approx \frac{\pi\alpha^2}{m_\mu^2}\left(1+\cos^2\theta\right),
\end{align}
and
\begin{equation}
    \sigma_{\text{C}} \approx \frac{8\pi\alpha^2}{3m_\mu^2},
\end{equation}
where $\alpha$ represents the electromagnetic coupling constant. 

Assume the mass of the new $U(1)$ gauge boson $Z'$ is much smaller than the mass of muon, $m_{Z'}\ll m_\mu$, one can immediately obtain the cross-section of the semi-Compton process, $\sigma_{\text{sC}}$ is simply Eq. \eqref{eq.sCsigma} with one of the coupling constant replaced by $\alpha\rightarrow\alpha'\equiv g_{Z'}^2/(4\pi)$, and it is given by
\begin{align}
	\sigma_{\text{sC}} &\approx \frac{8\pi\alpha\alpha'}{3m_\mu^2}\times\sqrt{1-\frac{m_{Z'}^2}{\omega^2}} ,\label{eq.sCsigma2}
\end{align}
in which we have multiplied an extra speed factor of the dark boson $v_{Z'}=\sqrt{1-m_{Z'}^2/\omega^2}$ to compensate for oversimplifying the cross-section $\sigma_{\text{sC}}$. The speed factor comes from the phase space integration of the final state particle $Z'$
\begin{align}
	\int d^3 k' &= 4\pi \int k'^{2} dk' \nonumber\\
	&=  4\pi \int (\omega^2 - m^2) d\omega \nonumber\\
	&= \int \sqrt{1-\frac{m_{Z'}^2}{\omega^2}} 4\pi\omega^2 d\omega.
\end{align}
Note that only $4\pi\omega^2$ appears in the integrand if the final state particle is a photon. The production rate is therefore given by
\begin{align}
	\Gamma_{\text{sC}}^{\text{prod}} &= n_{\mu}n_{\gamma}v_{\mu\gamma}\sigma_{\text{sC}} \\
	&= n_{\mu}\int\frac{d^3p_\gamma}{(2\pi)^3}\frac{1}{e^{E_{\gamma}/T}-1}\sigma_{\text{sC}}.
\end{align}
Since the SN neutrino energies are mostly distributed in a few tenths of MeV, we are safe to make an assumption that the energy of incident photons is actually much smaller than the mass of muons, i.e. $E_\gamma\ll m_\mu$. Consequently, the final state muons almost don't recoil and the energy of muon doesn't change after the scattering. Therefore, the energy of the initial state photon approximately equals the energy of the final state muon $E_\gamma\approx\omega$ due to the energy conservation law. The phase space integral in $n_\gamma$ can be integrated out with the energy conservation delta function, and the Boltzmann distribution function in the integrand is moved to the later $Z'$ phase space integral, resulting
\begin{align}
	\Gamma_{\text{sC}}^{\text{prod}} &\approx n_{\mu}\frac{1}{e^{\omega/T}-1}\sigma_{\text{sC}} \\
	&\approx \frac{8\pi\alpha\alpha'}{3m_\mu^2}\frac{n_{\mu}}{e^{\omega/T}-1}\sqrt{1-\frac{m_{Z'}^2}{\omega^2}}.
\end{align}
Finally, we apply a simple correction to account for the Pauli-blocking factor of muons by multiplying the above formula with a degeneracy factor $F_{\text{deg}}$. Therefore,
\begin{equation} \label{eq.sCproduction_a}
	\Gamma_{\text{sC}}^{\text{prod}}	\approx \frac{8\pi\alpha\alpha'}{3m_\mu^2}\frac{n_{\mu}F_{\text{deg}}}{e^{\omega/T}-1}\sqrt{1-\frac{m_{Z'}^2}{\omega^2}} ,
\end{equation}
where $F_{\text{deg}}$ is defined as
\begin{align}\label{eq.Fdeg_a}
	F_{\text{deg}} &\equiv \frac{1}{n_{\mu}^{\text{th}}}\int\frac{2d^3p}{(2\pi)^3}\frac{1}{e^{(E-\mu_\mu)/T}+1}\left(1-\frac{1}{e^{(E-\mu_\mu)/T}+1}\right),
\end{align}
in which
\begin{align}
	n_\mu &= \int \frac{2d^3p}{(2\pi)^3} \left( \frac{1}{e^{(E-\mu_\mu)/T}+1} - \frac{1}{e^{(E+\mu_\mu)/T}+1} \right),
 \end{align}
 and
 \begin{align}
	n_\mu^{\text{th}} &= \int \frac{2d^3p}{(2\pi)^3}\frac{1}{e^{(E-\mu_\mu)/T}+1},
\end{align}
in which the prefactor $2$ in the phase space integrand represents the spin degree of freedom of muons.
$n_\mu$ and $n_\mu^{\text{th}}$ are the net muon and thermal muon number density respectively. The idea of $F_{\text{deg}}$ is simply inserting the Pauli-blocking factor back into the integrand of $n_{\mu}^{\text{th}}$, this method is applicable only under the assumption that the final state muon does not recoil.
 Eq.~\eqref{eq.sCproduction_a} can also be derived by using its inverse process together with the principle of detail balance, the speed factor $v_{Z'}=\sqrt{1-m_{Z'}^2/\omega^2}$ will show up clearly during the derivation also.

Instead of applying the Klein-Nishina formula directly, a more sophisticated and complete calculation can be done, with some assumptions, from the general formula of the event rate phase space integral. In this way, we will be able to realize where the degeneracy term $F_{\text{deg}}$ comes from.

\noindent
\begin{minipage}{\columnwidth} %requires mpost command
	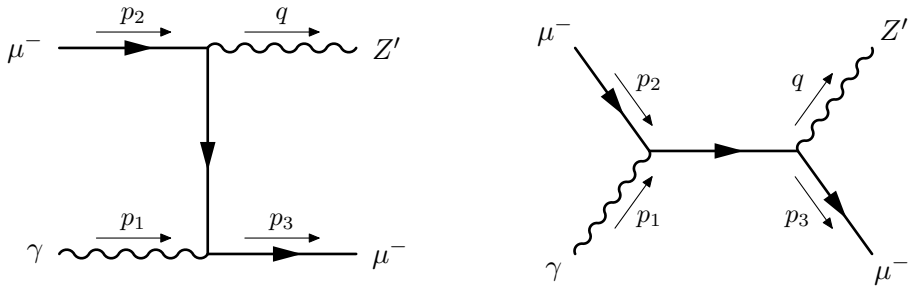
\begin{figure}[H]
		\centering
		\begin{fmffile}{semicomptonwithlabel_t}
			\setlength{\unitlength}{.3333\columnwidth}
			\vspace{1em}
			\begin{fmfgraph*}(0.9,0.5)
				\fmfpen{1.0}
				\fmfleft{i1,i2}
				\fmfright{o1,o2}
				\fmf{fermion}{i2,v2}
				\fmf{fermion}{v1,o1}
				\fmf{photon}{v2,o2}
				\fmf{photon}{i1,v1}
				\fmfv{l=$\mu^-$, l.a=180}{i2}
				\fmfv{l=$\gamma$, l.a=180}{i1}
				\fmfv{l=$Z'$, l.a=0}{o2}
				\fmfv{l=$\mu^-$, l.a=0}{o1} 
				\fmf{fermion,label=,tension=0}{v2,v1} 
				\marrow{a}{up}{top}{$p_1$}{i1,v1}
				\marrow{b}{up}{top}{$p_2$}{i2,v2}
				\marrow{c}{up}{top}{$p_3$}{v1,o1}
				\marrow{d}{up}{top}{$q$}{v2,o2}
			\end{fmfgraph*}%
		\end{fmffile}%
		\qquad\qquad
		\begin{fmffile}{semicomptonwithlabel_s}
			\setlength{\unitlength}{.3333\columnwidth}
			\vspace{1em}
			\begin{fmfgraph*}(0.9,0.5)
				\fmfpen{1.0}
				\fmfleft{i1,i2}
				\fmfright{o1,o2}
				\fmf{fermion}{i2,v1}
				\fmf{fermion}{v2,o1}
				\fmf{photon}{v2,o2}
				\fmf{photon}{i1,v1}
				\fmfv{l=$\mu^-$, l.a=135,l.d=1}{i2}
				\fmfv{l=$\gamma$, r.a=135}{i1}
				\fmfv{l=$Z'$, l.a=45,l.d=4}{o2}
				\fmfv{l=$\mu^-$, r.a=45,l.d=1}{o1} 
				\fmf{fermion,label=,tension=1.0}{v1,v2} 
				\Marrow{a}{right}{bot}{$p_1$}{i1,v1}{8}{14}
				\Marrow{b}{right}{top}{$p_2$}{i2,v1}{8}{14}
				\Marrow{c}{left}{bot}{$p_3$}{v2,o1}{8}{14}
				\Marrow{d}{left}{top}{$q$}{v2,o2}{8}{14}
			\end{fmfgraph*}%
		\end{fmffile}%
		\caption{Feynman diagrams of $Z'$ production in semi-Compton process with momentum labels. t channel on the left and s channel on the right.}
		\label{fig.sCprodwithlabels_a}
	\end{figure}%
\end{minipage}
\vspace{1em}

\noindent The Feynman diagram with momentum labels is given in Fig.~\eqref{fig.sCprodwithlabels_a} for later calculation convenience. Momentum labels $p_1,p_2,p_3$ and $q$ are used to represent the 4-momentum of initial state particles $\gamma,\mu^-$ and final state particles $\mu^-,Z'$ respectively. The event rate formula is given by
\begin{align}
	d\Gamma_{\text{sC}}^{\text{prod}}
	&=  \frac{1}{2\omega} \lvert \mathcal{M}_{\text{sC}}^{\text{prod}} (s,t) \rvert^2 (2\pi)^4\delta^4(q+p_3-p_1-p_2) \nonumber\\ &\qquad\qquad\times \frac{d^3p_1}{2E_{1}(2\pi)^3} f_1(E_1) \frac{2d^3p_2}{2E_{2}(2\pi)^3} f_2(E_2) \frac{2d^3p_3}{2E_{3}(2\pi)^3} (1- f_3(E_3)),
\end{align}
in which the pre-factor 2 in front of the initial and final state muon phase space $d^3p_2$ and $d^3p_3$ represent the spin degree of freedoms of Dirac particles. Multiplying the phase space integral of the new gauge boson $Z'$ $(1+f_q(\omega))d^3q/(2\pi)^3$ on both sides. We obtain
\begin{align}
	&\frac{d^3q}{(2\pi)^3}(1+f_q(\omega))d\Gamma_{\text{sC}}^{\text{prod}} \nonumber\\
	&=  \lvert \mathcal{M}_{\text{sC}}^{\text{prod}} (s,t) \rvert^2 (2\pi)^4\delta^4(q+p_3-p_1-p_2) \nonumber\\ &\qquad\qquad\times \frac{d^3p_1}{2E_{1}(2\pi)^3} f_1(E_1) \frac{2d^3p_2}{2E_{2}(2\pi)^3} f_2(E_2) \frac{2d^3p_3}{2E_{3}(2\pi)^3} (1- f_3(E_3)) \frac{d^3q}{2\omega(2\pi)^3} (1+ f_q(\omega)) \nonumber\\
	&= \lvert \mathcal{M}_{\text{sC}}^{\text{prod}} (s,t) \rvert^2 (2\pi)^4\delta^4(q+p_3-p_1-p_2) \nonumber\\ &\qquad\qquad\times \frac{d^3p_1}{2E_{1}(2\pi)^3} f_1(E_1)(1+ f_q(\omega)) \frac{2d^3p_2}{2E_{2}(2\pi)^3} f_2(E_2)(1- f_3(E_3)) \frac{2d^3p_3}{2E_{3}(2\pi)^3} \frac{d^3q}{2\omega(2\pi)^3}.
\end{align}
Since the temperature of a SN is around a few tenths of MeV, it can assume that the energy of the incident photon $\gamma$ is much less than the mass of muon $m_\mu\approx 105$ MeV. In other words, the muon almost doesn't recoil after the collision $E_3 \approx E_2$, therefore, $f_3(E_3)\approx f_2(E_2)$. Furthermore, because the process is considered to be in thermal equilibrium, we have $f_q(\omega) \approx f_1(E_1)$ by the energy conservation law $\omega\approx E_1$. With these two assumptions, the final state Pauli blocking as well as the Bose enhancement factors can be moved and combined with the initial state phase space integrals to give
\begin{align}
	&\frac{d^3q}{(2\pi)^3}(1+f_q(\omega))d\Gamma_{\text{sC}}^{\text{prod}} \nonumber\\
	&\approx \lvert \mathcal{M}_{\text{sC}}^{\text{prod}} (s,t) \rvert^2 (2\pi)^4\delta^4(q+p_3-p_1-p_2) \nonumber\\ &\qquad\qquad\times \frac{d^3p_1}{2E_{1}(2\pi)^3} f_1(E_1)(1+ f_1(E_1)) \frac{2d^3p_2}{2E_{2}(2\pi)^3} f_2(E_2)(1- f_2(E_2)) \frac{2d^3p_3}{2E_{3}(2\pi)^3} \frac{d^3q}{2\omega(2\pi)^3} \nonumber\\
	&= \frac{d^3p_1}{(2\pi)^3} f_1(E_1)(1+ f_1(E_1)) \frac{2d^3p_2}{(2\pi)^3} f_2(E_2)(1- f_2(E_2))  \nonumber\\ &\qquad\qquad\times \frac{1}{2E_{1}2E_{2}} \lvert \mathcal{M}_{\text{sC}}^{\text{prod}} (s,t) \rvert^2 (2\pi)^4\delta^4(q+p_3-p_1-p_2) \frac{2d^3p_3}{2E_{3}(2\pi)^3} \frac{d^3q}{2\omega(2\pi)^3}.
\end{align}
By replacing one $\alpha$ to $\alpha'$ in $\lvert \mathcal{M}_{\text{sC}}^{\text{prod}}\rvert$ we get
\begin{align}
	&\frac{d^3q}{(2\pi)^3}(1+f_q(\omega))d\Gamma_{\text{sC}}^{\text{prod}} \nonumber\\
	&= \frac{d^3p_1}{(2\pi)^3} f_1(E_1)(1+ f_1(E_1)) \times \frac{2d^3p_2}{(2\pi)^3} f_2(E_2)(1- f_2(E_2)) \times \sqrt{1-\frac{m_{Z'}^2}{\omega^2}}\sigma_{\text{C}}(s) \nonumber\\
	&\approx \frac{d^3q}{(2\pi)^3} f_q(\omega)(1+ f_q(\omega)) \times n_{\mu}F_{deg}\times \sqrt{1-\frac{m_{Z'}^2}{\omega^2}} \times \frac{8\pi\alpha\alpha'}{3m_\mu^2},
\end{align}
in which the degeneracy factor $F_{\text{deg}}$ is denoted in green color. One can verify that it is just Eq.~\eqref{eq.Fdeg_a}. Finally, the $Z'$ production rate in semi-Compton process $\Gamma_{\text{sC}}^{\text{prod}}$ can be obtained by dividing $(1+f_q(\omega))d^3q/(2\pi)^3$ on both sides. Results in
\begin{equation}
		\Gamma_{\text{sC}}^{\text{prod}}	\approx \frac{8\pi\alpha\alpha'}{3m_\mu^2}\frac{n_{\mu}F_{\text{deg}}}{e^{\omega/T}-1}\sqrt{1-\frac{m_{Z'}^2}{\omega^2}} 
\end{equation}
Therefore, the direct calculation from the general formula reveal the form of Eq.~\eqref{eq.sCproduction_a} under the assumptions $f_3(E_3)\approx f_2(E_2)$ and $f_q(\omega) \approx f_1(E_1)$ as expected.

\begin{comment}
	A more sophisticated and complete calculation can be done by using a computer program, the full cross-section of semi-Compton tree-level diagram $\mu^- +\gamma\rightarrow\mu^- +Z'$ without any approximation is
	\begin{align}
		\sigma_{\text{sC}} &= -\frac{\alpha g_{Z'}^2}{4 \left(2 m_{\mu} \omega+m_{Z'}^2\right)^3 \left(m_{\mu}^2+2 m_{\mu} \omega+m_{Z'}^2\right)^2} \nonumber\\
		&\times \left[ 2 \left(m_{\mu}^2+2 m_{\mu} \omega+m_{Z'}^2\right)^2 \left(4 m_{\mu}^2 \left(\omega^2-m_{Z'}^2\right) -8 m_{\mu}^3 \omega-8 m_{\mu}^4+m_{Z'}^4\right) \right.\nonumber\\
		&\left. \times\log \left(\frac{m_{\mu}-\sqrt{\omega^2-m_{Z'}^2}+\omega}{m_{\mu}+\sqrt{\omega^2-m_{Z'}^2}+\omega}\right)-4 m_{\mu} \sqrt{\omega^2-m_{Z'}^2} \left(4 m_{\mu}^4 \left(5 m_{Z'}^2+9 \omega^2\right) \right.\right. \nonumber\\
		&\left.\left. + 4 m_{\mu}^3 \left(13 m_{Z'}^2 \omega+\omega^3\right)+m_{\mu}^2 \left(20 m_{Z'}^2 \omega^2+17 m_{Z'}^4\right)\right.\right. \nonumber\\
		&\left.\left. +32 m_{\mu}^5 \omega+8 m_{\mu}^6+17 m_{\mu} m_{Z'}^4 \omega+4 m_{Z'}^6\right)\right] \label{eq.fullsCcross}
	\end{align}
	One can verify that under the assumption $\omega\ll m_\mu$, Eq. \eqref{eq.fullsCcross} can be simplified to the Klein-Nishina formula Eq. \eqref{eq.sCsigma2} with $g_{Z'} \equiv 4\pi \alpha'$.
\end{comment}

%InvSemiComp
The $Z'$ absorption rate through semi-Compton process, as indicated in Fig.~\ref{fig.sCabswithlabels_a}, can be obtained by applying the principle of detailed balance on the aforementioned $Z'$ production rate Eq.~\eqref{eq.sCproduction_a}, giving
\begin{align}
	\Gamma_{\text{sC}}^{\text{abs}} &= \frac{2}{3} e^{\omega/T}\times\Gamma_{\text{sC}}^{\text{prod}} \nonumber\\
	&\approx \frac{2}{3} e^{\omega/T} \times \frac{8\pi\alpha\alpha'}{3m_\mu^2}\frac{n_{\mu}F_{\text{deg}}}{e^{\omega/T}-1}\sqrt{1-\frac{m_{Z'}^2}{\omega^2}} .
\end{align}
Since a standard model photon $\gamma$ has only two polarization while $Z'$ has three, the pre-factor $2/3$ accounts for the correct polarization averaging of the initial $Z'$ state in replacing the initial photon state. Results in
\begin{equation} \label{eq.sCabsorption_a}
	\Gamma_{\text{sC}}^{\text{abs}}
	\approx  \frac{16\pi\alpha\alpha'}{9m_\mu^2} n_{\mu} F_{\text{deg}} (1+\frac{1}{e^{\omega/T}-1}) \sqrt{1-\frac{m_{Z'}^2}{\omega^2}} ,
\end{equation}
which can be verified by a direct calculation from the event rate general formula under some assumptions.

\noindent
\begin{minipage}{\columnwidth} %requires mpost command
	\begin{figure}[H]
		\centering
		\begin{fmffile}{invsemicomptonwithlabel_t}
			\setlength{\unitlength}{.3333\columnwidth}
			\vspace{1em}
			\begin{fmfgraph*}(0.9,0.5)
				\fmfpen{1.0}
				\fmfleft{i1,i2}
				\fmfright{o1,o2}
				\fmf{photon}{i2,v2}
				\fmf{photon}{v1,o1}
				\fmf{fermion}{v2,o2}
				\fmf{fermion}{i1,v1}
				\fmfv{l=$Z'$, l.a=180}{i2}
				\fmfv{l=$\mu^-$, l.a=180}{i1}
				\fmfv{l=$\mu^-$, l.a=0}{o2}
				\fmfv{l=$\gamma$, l.a=0}{o1} 
				\fmf{fermion,label=,tension=0}{v1,v2} 
				\marrow{a}{up}{top}{$p_2$}{i1,v1}
				\marrow{b}{up}{top}{$q$}{i2,v2}
				\marrow{c}{up}{top}{$p_4$}{v1,o1}
				\marrow{d}{up}{top}{$p_3$}{v2,o2}
			\end{fmfgraph*}%
		\end{fmffile}%
		\qquad\qquad
		\begin{fmffile}{invsemicomptonwithlabel_s}
			\setlength{\unitlength}{.3333\columnwidth}
			\vspace{1em}
			\begin{fmfgraph*}(0.9,0.5)
				\fmfpen{1.0}
				\fmfleft{i1,i2}
				\fmfright{o1,o2}
				\fmf{fermion}{i1,v1}
				\fmf{fermion}{v2,o2}
				\fmf{photon}{v2,o1}
				\fmf{photon}{i2,v1}
				\fmfv{l=$\mu^-$, r.a=135,l.d=1}{i1}
				\fmfv{l=$Z'$, r.a=135,l.d=4}{i2}
				\fmfv{l=$\gamma$, r.a=45,l.d=4}{o1}
				\fmfv{l=$\mu^-$, r.a=45,l.d=4}{o2} 
				\fmf{fermion,label=,tension=1.0}{v1,v2} 
				\Marrow{a}{right}{bot}{$p_2$}{i1,v1}{8}{14}
				\Marrow{b}{right}{top}{$q$}{i2,v1}{8}{14}
				\Marrow{c}{left}{bot}{$p_4$}{v2,o1}{8}{14}
				\Marrow{d}{left}{top}{$p_3$}{v2,o2}{8}{14}
			\end{fmfgraph*}%
		\end{fmffile}%
		\caption{The Feynman diagrams of $Z'$ absorption in semi-Compton process with momentum labels. t channel on the left and s channel on the right.}
		\label{fig.sCabswithlabels_a}
	\end{figure}
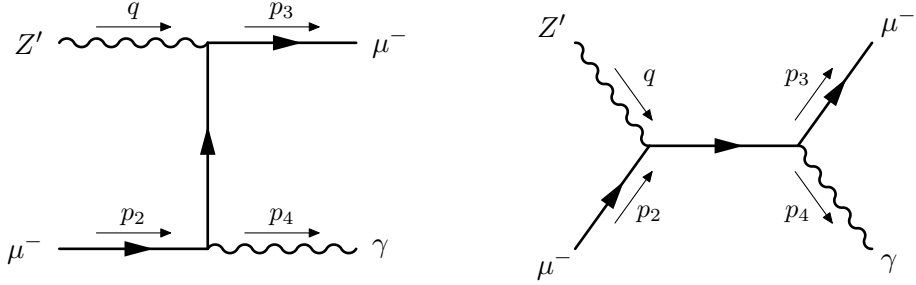%
\end{minipage}%
\vspace{1em}

The absorption rate formula of $Z'$ through the semi-Compton process is given by
\begin{align}
	d\Gamma_{\text{sC}}^{\text{abs}} &=  \frac{1}{2\omega}\lvert \mathcal{M}_{\text{sC}}^{\text{abs}} (s,t) \rvert^2 (2\pi)^4\delta^4(q+p_2-p_3-p_4) \nonumber\\ &\qquad\times  \frac{2d^3p_2}{2E_{2}(2\pi)^3} f_2(E_2) \frac{d^3p_3}{2E_{3}(2\pi)^3} (1- f_3(E_3)) \frac{d^3p_4}{2E_4(2\pi)^3} (1+ f_4(E_4))
\end{align}
Since the temperature of SNe is around a few tenths of MeV, which is much smaller than the mass of muon $m_\mu$, we assume the energy of the initial new gauge boson is much smaller than the mass of $muon$, $\omega \ll m_\mu$. In other words, the muon does not recoil under this assumption as the energy of the initial and final state muon does not change. Therefore, the energy of the final state photon approximately equals to the energy of the initial state $Z'$, i.e., $E_4 \approx \omega$. The Bose enhancement factor as well as the Pauli blocking factor of the final state muon $\mu$ and photon $\gamma$ can then be represented by the initial state muon and $Z'$ energies. Eventually, they can be factored out from the final state phase-space integral and combined with the initial state phase-space integral. The equation above becomes
\begin{align}
	d\Gamma_{\text{sC}}^{\text{abs}} &\approx \frac{1}{2\omega} \lvert \mathcal{M}_{\text{sC}}^{\text{abs}} (s,t) \rvert^2 (2\pi)^4\delta^4(q+p_2-p_3-p_4) \nonumber\\ &\qquad\times(1+ f_4(\omega)) \frac{2d^3p_2}{2E_{2}(2\pi)^3} f_2(E_2)(1- f_2(E_2)) \frac{d^3p_3}{2E_{3}(2\pi)^3} \frac{d^3p_4}{2E_4(2\pi)^3} \\
	&= (1+ f_1(\omega)) \frac{2d^3p_2}{(2\pi)^3} f_2(E_2)(1- f_2(E_2))  \nonumber\\ &\qquad\times \frac{1}{2\omega 2E_{2}} \lvert \mathcal{M}_{\text{sC}}^{\text{abs}} (s,t) \rvert^2 (2\pi)^4\delta^4(q+p_2-p3-p4) \frac{d^3p_3}{2E_{3}(2\pi)^3} \frac{d^3p_4}{2E_4(2\pi)^3}.
\end{align}
Since the semi-Compton process dominates the constraint only in small $m_{Z'}$, in this region, the amplitude square $\lvert \mathcal{M}_{\text{sC}}^{\text{abs}}(s,t) \rvert^2 \approx (2/3)\times\lvert \mathcal{M}_{\text{C}}(s,t) \rvert^2$. The pre-factor $2/3$ accounts for the correct polarization averaging in replacing $\gamma$ by $Z'$. Therefore,
\begin{align}
	d\Gamma_{\text{sC}}^{\text{abs}} &\approx (1+ f_1(\omega)) \times \frac{2d^3p_2}{(2\pi)^3} f_2(E_2)(1- f_2(E_2)) \times \frac{2}{3} \lvert v_1-v_2\rvert \sigma_{\text{C}}.
 \end{align}
Consequently,
 \begin{align}
	\Gamma_{\text{sC}}^{\text{abs}} &=  \frac{2}{3}(1+ f_1(\omega)) \times n_{\mu}F_{deg} \times \sqrt{1-\frac{m_{Z'}^2}{\omega^2}}\sigma_{\text{C}}(\alpha^2\rightarrow\alpha\alpha_{Z'}),
\end{align}
where the distribution $f_1(\omega)$ is the Bose-Einstein distribution and $\sigma_{\text{C}}(\alpha^2\rightarrow\alpha\alpha_{Z'}) \approx 8\pi \alpha\alpha_{Z'}/(3m_\mu^2)$. Finally, the $Z'$ absorption rate through a semi-Compton process is 
\begin{equation}
	\Gamma_{\text{sC}}^{\text{abs}} \approx  \frac{ 16\pi \alpha\alpha_{Z'}}{9m_\mu^2}(1+ \frac{1}{e^{\omega/T}-1})  n_{\mu}F_{deg}  \sqrt{1-\frac{m_{Z'}^2}{\omega^2}}.
\end{equation}

\section{Loop Bremsstrahlung}\label{sec.LoopBrem}

\begin{figure}[b!]
    \centering
    \begin{minipage}{.5\columnwidth} %requires mpost command
		\begin{figure}[H]
			\centering
			\begin{fmffile}{brem}
				\setlength{\unitlength}{.6\columnwidth}
				\vspace{1em}
				\begin{fmfgraph*}(0.9,0.5)
					\fmfstraight
					\fmfpen{1.0}
					\fmfleft{i1,i2,i3}
					\fmfright{o1,o2,o3}
					\fmf{fermion}{i2,v2}
					\fmf{fermion}{v1,o1}
					\fmf{fermion}{v2,o2}
					\fmf{fermion}{i1,v1}
					\fmffreeze
					%\fmf{phantom}{o1,o3,o2}
					\fmf{phantom}{v2,v3,h1,o2}
					\fmffreeze
					\fmf{photon}{v3,o3}
					\fmfv{l=$p$, l.a=180}{i2}
					\fmfv{l=$n$, l.a=180}{i1}
					\fmfv{l=$p$, l.a=0}{o2}
					\fmfv{l=$n$, l.a=0}{o1} 
					\fmfv{l=$\gamma$, l.a=0}{o3}
					\fmf{dashes,label=,tension=0}{v2,v1} 
					\marrow{a}{up}{top}{$p_2$}{i1,v1}
					\marrow{b}{up}{top}{$p_1$}{i2,v2}
					\marrow{c}{up}{top}{$p_4$}{v1,o1}
					\marrow{d}{up}{top}{$p_3$}{h1,o2}
					\Marrow{e}{up}{lft}{$k$}{v3,o3}{8}{15}
				\end{fmfgraph*}%
			\end{fmffile}%
			\caption*{Bremsstrahlung}
		\end{figure}%
	\end{minipage}%
\begin{minipage}{.5\columnwidth} %requires mpost command
		\begin{figure}[H]
			\centering
			\begin{fmffile}{loop}
				\setlength{\unitlength}{.6\columnwidth}
				\vspace{1em}
				\begin{fmfgraph*}(0.9,0.5)
					\fmfpen{1.0}
					\fmfleft{i1}
					\fmfright{o1}
					\fmfkeep{fermion}
					\fmf{phantom}{i1,v1,v2,o1}
					\fmf{photon}{i1,v1}
					\fmf{photon}{v2,o1}
					\fmf{fermion,left,label.side=left,tension=.6}{v1,v2}
					\fmf{phantom,left,label=$\mu,,\tau$,label.side=right,label.dist=16,tension=0}{v1,v2}
					\fmf{fermion,left,tension=.6}{v2,v1}
					\fmfv{l=$\gamma$, l.a=180}{i1}
					\fmfv{l=$Z^{'}$, l.a=0}{o1} 
					\marrow{a}{up}{top}{$k$}{i1,v1}
					\marrow{b}{up}{top}{$k$}{v2,o1}
					\Marrow{c}{up}{top}{$k+p$}{v1,v2}{26}{26}
					\Marrow{d}{down}{bot}{$p$}{v2,v1}{26}{26}
				\end{fmfgraph*}%
			\end{fmffile}%
			\caption*{Loop}
		\end{figure}%
	\end{minipage}%
    \caption{Feynman diagrams indicating $Z'$ production through $n-p$ loop Bremsstrahlung.}
    \label{fig:Brem_Loop}
\end{figure}
Feynman diagrams for $Z'$ production from loop Bremsstrahlung are given in Fig.~\ref{fig:Brem_Loop}.
The emission rate of $Z'$ from $n-p$ loop Bremsstrahlung is related to the matrix element by
\begin{align}
	d\Gamma_{\text{brem}} &= \frac{1}{2\omega} \frac{d^3 p_1 f(p_1)}{(2\pi)^3 2E_1} \frac{d^3 p_2 f(p_2)}{(2\pi)^3 2E_2} \frac{d^3 p_3(1-f(p_3))}{(2\pi)^3 2E_3} \frac{d^3 p_4 (1-f(p_4))}{(2\pi)^3 2E_4} \nonumber\\
	&\hspace{15em} \times (2\pi)^4\delta^4(p_1+p_2-p_3-p_4-k) |\mathcal{M}|^2_{np\gamma} \nonumber\\
	&\approx \frac{1}{2\omega} \frac{d^3 p_1 f(p_1)}{(2\pi)^3 2E_1} \frac{d^3 p_2 f(p_2)}{(2\pi)^3 2E_2} \frac{d^3 p_3}{(2\pi)^3 2E_3} \frac{d^3 p_4 }{(2\pi)^3 2E_4} \nonumber\\
	&\hspace{15em} \times (2\pi)^4\delta^4(p_1+p_2-p_3-p_4-k) |\mathcal{M}|^2_{np\gamma}
\end{align}
It can be written in our shorthand notation for later calculation convenience
\begin{align}
	d\Gamma_{\text{brem}} &= \frac{1}{2\omega} d_{\text{LIPS}}^{i,2} d_{\text{LIPS}}^{f,2} \Delta^4_5 |\mathcal{M}|^2_{np\gamma} \\
	d\Gamma_{\text{brem}} d_{\text{LIPS}}^{\gamma}&= \frac{1}{2\omega} d_{\text{LIPS}}^{i,2} d_{\text{LIPS}}^{f,3} \Delta^4_5 |\mathcal{M}|^2_{np\gamma} \nonumber\\
	&= \frac{1}{2\omega} d_{\text{LIPS}}^{i,2} d\sigma_{np\gamma}F,
\end{align}
where $F= 2E_1 E_2 v_{\text{rel}}$, $d_{\text{LIPS}}^{i,2}$ and $d_{\text{LIPS}}^{f,2}$ are the initial and final state Lorentz invariant phase space integral, and $\Delta^4_5 \equiv (2\pi)^4\delta^4(p_1+p_2-p_3-p_4-k)$.
With the Soft Radiation Approximation (SRA)~\cite{Low:1958sn,Rrapaj:2015wgs}, the cross section of Bremsstrahlung $d\sigma_{np\gamma}$ can be expressed in $n-p$ elastic collision cross-section $d\sigma_{np}$ by
\begin{align}
	d\sigma_{np\gamma} &= 4\pi \alpha \epsilon^2 d_{\text{LIPS}}^\gamma (\epsilon^\mu J_\mu)^2 d\sigma_{np}.
\end{align}
Expressing $d\sigma_{np\gamma}$ in $d\sigma_{np}$, the emission rate is 
\begin{align}
	d\Gamma_{\text{brem}}^{\text{SRA}} &=  \frac{1}{2\omega} d_{\text{LIPS}}^{i,2} d_{\text{LIPS}}^{f,2} (4\pi \alpha \epsilon^2 d_{\text{LIPS}}^\gamma (\epsilon^\mu J_\mu)^2  \Delta^4_4 |\mathcal{M}|^2_{np}) \\
	&=  \frac{4\pi\alpha\epsilon^2}{2\omega} d_{\text{LIPS}}^{i,2} d_{\text{LIPS}}^{f,2}(\epsilon^\mu J_\mu)^2 \Delta^4_4 64\pi^2 E_{\text{cm}}^2 \frac{d\sigma_{np}}{d\Omega_{\text{cm}}} \nonumber\\
	&=  \frac{2\pi}{\omega} \alpha\epsilon^2 d_{\text{LIPS}}^{i,2} d_{\text{LIPS}}^{f,2}(\epsilon^\mu J_\mu)^2 \Delta^4_4 32\pi E_{\text{cm}}^2 \frac{d\sigma_{np}}{d\theta_{\text{cm}}}.
\end{align}
In which, $E_{\rm cm}$ and $\Omega_{\rm cm}$ denote the total energy of the initial particles and the solid angle in the center-of-momentum frame, respectively. The differential cross-section $\frac{d\sigma_{np}}{d\theta_{\text{cm}}}$ is calculated based on the Reid93 nucleon-nucleon potential~\cite{Zhaba:2018iyj} (the result can be found at https://nn-online.org). Rewriting it in the usual notation, it becomes
\begin{align}
	d\Gamma_{\text{brem}}^{\text{SRA}} &= \frac{2\pi}{\omega} \alpha\epsilon^2  \frac{d^3 p_1 f(p_1)}{(2\pi)^3 2E_1} \frac{d^3 p_2 f(p_2)}{(2\pi)^3 2E_2} \frac{d^3 p_3}{(2\pi)^3 2E_3} \frac{d^3 p_4}{(2\pi)^3 2E_4}  \nonumber\\
	&\hspace{10em} \times (\epsilon^\mu J_\mu)^2 (2\pi)^4\delta^4(p_1+p_2-p_3-p_4) 32\pi E_{\text{cm}}^2 \frac{d\sigma_{np}}{d\theta_{\text{cm}}},
\end{align}
where $\delta^4(p_1+p_2-p_3-p_4)=\delta(E_{\text{cm}}-E_3-E_4)\delta^3(\vec{p}_3+\vec{p}_4)$ and the energy $E=\sqrt{M^2+\lvert \vec{p}\rvert^2}\approx M+\lvert \vec{p}\rvert^2/(2M)$ in non-relativistic limit. The three-dimensional delta function ensures $E_3 = E_4$. Therefore $E_1=E_2=E_3=E_4=E_{\text{cm}}/2$. The delta function can be rewritten in
\begin{align}
	\delta^4(p_1+p_2-p_3-p_4) &= \delta(T_{\text{cm}} - \frac{\lvert \vec{p}_3 \rvert^2}{M}) \delta^3(\vec{p}_3+\vec{p}_4)\nonumber\\
	&= \frac{M}{2\lvert \vec{p}_3\rvert}\delta( \lvert \vec{p}_3\rvert -\sqrt{MT_{\text{cm}}} )\delta^3(\vec{p}_3+\vec{p}_4)
\end{align}
Where $T_{\rm cm}=E_{\rm cm}-2M$. By expressing $E_i$ in terms of $E_{\text{cm}}$ and substituting the 4-dimensional delta function with the aforementioned result, the integration of $p_4$ can be carried out, leading to the determination of the emission rate as
\allowdisplaybreaks
\begin{align}
	d\Gamma_{\text{brem}}^{\text{SRA}} &=  \frac{\alpha\epsilon^2}{2\pi\omega E_{\text{cm}}^2}   \frac{d^3 p_1 f(p_1)}{(2\pi)^3 E_{\text{cm}}} \frac{d^3 p_2 f(p_2)}{(2\pi)^3 E_{\text{cm}}} d^3 p_3  \nonumber\\
	&\hspace{5em} \times (\epsilon^\mu J_\mu)^2 \frac{M}{2\lvert \vec{p}_3\rvert}\delta( \lvert \vec{p}_3\rvert -\sqrt{MT_{\text{cm}}} ) 32\pi E_{\text{cm}}^2 \frac{d\sigma_{np}}{d\theta_{\text{cm}}} \nonumber\\
	&=  \frac{\alpha\epsilon^2}{2\pi\omega}   \frac{d^3 p_1 f(p_1)}{(2\pi)^3 E_{\text{cm}}} \frac{d^3 p_2 f(p_2)}{(2\pi)^3 E_{\text{cm}}}  p_3^2dp_3  \nonumber\\
	&\hspace{5em} \times d\Omega_3(\epsilon^\mu J_\mu)^2 \frac{M}{2\lvert \vec{p}_3\rvert}\delta( \lvert \vec{p}_3\rvert -\sqrt{MT_{\text{cm}}} ) 32\pi \frac{d\sigma_{np}}{d\theta_{\text{cm}}} \nonumber\\
	&=  \frac{\alpha\epsilon^2 M^{3/2}}{4\pi^5 \omega}   \frac{d^3 p_1 f(p_1)}{ E_{\text{cm}}} \frac{d^3 p_2 f(p_2)}{E_{\text{cm}}}  d\theta_{\text{cm}} (\epsilon^\mu J_\mu)^2 \sqrt{T_{\text{cm}}}  \frac{d\sigma_{np}}{d\theta_{\text{cm}}} \\
	&\approx  \frac{\alpha\epsilon^2 }{16\pi^5 \omega M^{1/2}}  d^3 p_1 f(p_1) d^3 p_2 f(p_2)  d\theta_{\text{cm}} (\epsilon^\mu J_\mu)^2 \sqrt{T_{\text{cm}}}  \frac{d\sigma_{np}}{d\theta_{\text{cm}}}.
\end{align}
Since we are working in a non-relativistic limit, we take $E_{\text{cm}}\approx 2M$ in the last equality.

Now, we take the Maxwell-Boltzmann distribution as the energy distribution of the initial nucleons such that
\begin{align}
	f(p)=\frac{n}{2}\left(\frac{2\pi}{MT}\right)^{3/2} e^{-\lvert \vec{p} \rvert^2/(2MT)}.
\end{align}
The emission rate can then be written as
\begin{align}
	d\Gamma_{\text{brem}}^{\text{SRA}} &=  \frac{n_n n_p \alpha\epsilon^2 }{8\pi^2 \omega M^{7/2}T^3} d^3 p_1  d^3 p_2 e^{-(\lvert \vec{p}_1 \rvert^2 + \lvert \vec{p}_2 \rvert^2)/(2MT)} (\epsilon^\mu J_\mu)^2 \sqrt{T_{\text{cm}}}  \frac{d\sigma_{np}}{d\theta_{\text{cm}}} d\theta_{\text{cm}}.
\end{align}
We define $\vec{p}_+ \equiv \vec{p}_1 + \vec{p}_2$ and $\vec{p}_- \equiv \vec{p}_1 - \vec{p}_2$ such that one of the momenta is perpendicular to $T_{\text{cm}}$ and can be integrated out immediately. Here $\lvert \vec{p}_- \rvert^2 = 4MT_{\text{cm}}$ and $\vec{p}_+$ is irrelevant to $T_{\text{cm}}$. We can immediately work out
\begin{align}
	d^3 p_1 d^3 p_2 &= \frac{d^3 p_+ d^3 p_-}{8} \\
	\lvert\vec{p}_1\rvert^2 + \lvert\vec{p}_2\rvert^2 &= \frac{\lvert\vec{p}_+\rvert^2 + \lvert\vec{p}_-\rvert^2}{2}.
\end{align}
Since $\vec{p}_+$ is irrelevant to $T_{\text{cm}}$, $d^3p_+$ can be integrated out by using the Gaussian integral, giving
\begin{align}
	\int_{0}^{\infty}\lvert\vec{p}_+\rvert^{2}e^{-\lvert\vec{p}_+\rvert^2/(4MT)} dp_+ &= (MT)^{3/2} \sqrt{4\pi}.
\end{align}
Therefore, the emission rate in the new coordinate system is
\begin{align}
	d\Gamma_{\text{brem}}^{\text{SRA}} &=  \frac{n_n n_p \alpha\epsilon^2 }{64\pi^2 \omega M^{7/2}T^3} d^3 p_+  d^3 p_- e^{-(\lvert \vec{p}_+ \rvert^2 + \lvert \vec{p}_- \rvert^2)/(4MT)} (\epsilon^\mu J_\mu)^2 \sqrt{T_{\text{cm}}}  \frac{d\sigma_{np}}{d\theta_{\text{cm}}} d\theta_{\text{cm}} \nonumber\\
	&=  \frac{n_n n_p \alpha\epsilon^2 \sqrt{4\pi}}{16\pi \omega M^2 T^{3/2}}  d^3 p_- e^{-\lvert \vec{p}_- \rvert^2/(4MT)} (\epsilon^\mu J_\mu)^2 \sqrt{T_{\text{cm}}}  \frac{d\sigma_{np}}{d\theta_{\text{cm}}} d\theta_{\text{cm}} \nonumber\\
	&=  \frac{n_n n_p \alpha\epsilon^2 \sqrt{4\pi}}{4 \omega M^2 T^{3/2}}  \lvert \vec{p}_- \rvert^2 dp_- e^{-\lvert \vec{p}_- \rvert^2/(4MT)} (\epsilon^\mu J_\mu)^2 \sqrt{T_{\text{cm}}}  \frac{d\sigma_{np}}{d\theta_{\text{cm}}} d\theta_{\text{cm}} \nonumber\\
	&=  \frac{n_n n_p \alpha\epsilon^2 \sqrt{4\pi} }{\omega M T^{3/2}} dp_- e^{-\lvert \vec{p}_- \rvert^2/(4MT)} (\epsilon^\mu J_\mu)^2 (T_{\text{cm}})^{3/2}  \frac{d\sigma_{np}}{d\theta_{\text{cm}}} d\theta_{\text{cm}} \nonumber\\
	&=  \frac{n_n n_p \alpha\epsilon^2 \sqrt{4\pi} }{\omega \sqrt{M} T^{3/2}} dT_{\text{cm}} e^{-T_{\text{cm}}/T} (\epsilon^\mu J_\mu)^2 T_{\text{cm}}  \frac{d\sigma_{np}}{d\theta_{\text{cm}}} d\theta_{\text{cm}},
\end{align}
in which the polarization sum of the massive gauge boson Z' in $(\epsilon^{\mu}J_{\mu})^2$ in given by. 
\begin{align}
	(\epsilon^{\mu}J_{\mu})^2 &= \sum_{i} \epsilon^{\mu}_i \epsilon^{\nu}_i J_{\mu} J_{\nu} \nonumber\\
	&= \left( -g^{\mu\nu}+\frac{k^\mu k^\nu}{m_{Z'}} \right) J_{\mu} J_{\nu}.
\end{align}
Recall that
\begin{align}
	J^{\mu} &= \left( \frac{p_1^\mu}{p_1 k} - \frac{p_2^\mu}{p_2 k} \right).
\end{align}
Therefore, the second term in the parenthesis vanishes. Such that
\begin{align}
	(\epsilon^{\mu}J_{\mu})^2 &= -g^{\mu\nu}J_\mu J_\nu = -J^2 \nonumber\\
	&= -\left( \frac{M^2}{p_1 k} + \frac{M^2}{p_2 k} -  \frac{p_1 p_2}{(p_1 k)(p_2 k)} \right).
\end{align}
Since we're in non-relativistic limit $(\lvert\vec{p}\rvert\ll M)$, to $\mathcal{O}(1/M^2)$,
\begin{align}
	(\epsilon^{\mu}J_{\mu})^2 &\approx \frac{1}{\omega^2}\left[ \frac{(\vec{p}_1 - \vec{p}_3)^2}{M^2} - \left( \frac{(\vec{p}_1 - \vec{p}_3)\cdot \vec{k}}{M \omega} \right)^2 \right].
\end{align}
In this simple form, the average over all $Z'$ emission angle can be done easily
\begin{align}
	\frac{1}{4\pi}\int d\Omega_{Z'} (\epsilon^{\mu}J_{\mu})^2 &\approx \frac{1}{4\pi}\int d\Omega_{Z'} \frac{1}{\omega^2}\left[ \frac{(\vec{p}_1 - \vec{p}_3)^2}{M^2} - \left( \frac{(\vec{p}_1 - \vec{p}_3)\cdot \vec{k}}{M \omega} \right)^2 \right] \nonumber\\
	&=\frac{(\vec{p}_1-\vec{p}_3)^2}{\omega^2 M^2}\left(1-\frac{\lvert\vec{k}\rvert^2}{3\omega^2}\right) \nonumber\\
	&=\frac{\lvert\vec{p}_1\rvert^2+\lvert\vec{p}_2\rvert^2 - 2\lvert\vec{p}_1\rvert \lvert\vec{p}_2\rvert \cos\theta_{\text{cm}} }{\omega^2 M^2}\left(1-\frac{\lvert\vec{k}\rvert^2}{3\omega^2}\right) \\
	&\approx \frac{2\lvert\vec{p}_1\rvert^2}{\omega^2 M^2}\left(1-\frac{\lvert\vec{k}\rvert^2}{3\omega^2}\right)(1-\cos\theta_{\text{cm}}) \\%
	&=\frac{2T_{\text{cm}}}{\omega^2 M} \left(1-\frac{\lvert\vec{k}\rvert^{2}}{3\omega^{2}}\right) (1-\cos\theta_{\text{cm}}).
\end{align}
In the soft limit $\lvert\vec{p}_1\rvert \approx \lvert\vec{p}_2\rvert$, it has been applied to the last two equalities. Insert it into the emission rate $d\Gamma_{\text{brem}}$ result in 
\begin{align}
	d\Gamma_{\text{brem}}^{\text{SRA}} &=  \frac{n_n n_p \alpha\epsilon^2 \sqrt{4\pi} }{\omega \sqrt{M} T^{3/2}} dT_{\text{cm}} e^{-T_{\text{cm}}/T} (\epsilon^\mu J_\mu)^2 T_{\text{cm}}  \frac{d\sigma_{np}}{d\theta_{\text{cm}}} d\theta_{\text{cm}} \nonumber\\
	&=  \frac{4n_n n_p \alpha\epsilon^2 \sqrt{\pi} }{\omega^3 (M T)^{3/2}} dT_{\text{cm}} e^{-T_{\text{cm}}/T} T_{\text{cm}}^2 \left(1-\frac{\lvert\vec{k}\rvert^2}{3\omega^2}\right) \sigma^{(2)}_{np}(T_{\text{cm}}) \label{eq.bremevrate},
\end{align} 
where the definition of $\sigma^{(2)}_{np}(T_{\text{cm}})$ is
\begin{align}
	\sigma^{(2)}_{np}(T_{\text{cm}}) &\equiv \int_{-1}^{1} (1-\cos\theta_{\text{cm}})\frac{d\sigma_{np}}{d\cos\theta_{\text{cm}}} d\cos\theta_{\text{cm}}.
\end{align}
The emissivity can be calculated by
\begin{align}
	\dot{\epsilon} &\equiv \int \int \frac{d^3 k}{(2\pi)^3} \omega d\Gamma_{\text{brem}}^{\text{SRA}} \\
	&= \frac{4n_n n_p \alpha\epsilon^2 \sqrt{\pi} }{(M T)^{3/2}} \int\int \frac{d^3 k}{(2\pi)^3}  \frac{1}{\omega^2 }dT_{\text{cm}} e^{-T_{\text{cm}}/T} T_{\text{cm}}^2 \left(1-\frac{\lvert\vec{k}\rvert^2}{3\omega^2}\right) \sigma^{(2)}_{np}(T_{\text{cm}}) \\
	&=  \frac{2n_n n_p \alpha\epsilon^2 }{(\pi M T)^{3/2}} \int\int dk \frac{\lvert\vec{k}\rvert^2}{\omega^2 } \left(1-\frac{\lvert\vec{k}\rvert^2}{3\omega^2}\right) dT_{\text{cm}} e^{-T_{\text{cm}}/T} T_{\text{cm}}^2 \sigma^{(2)}_{np}(T_{\text{cm}}).
\end{align}
From the conservation of energy, the $\vec{k}$ phase space integral should be integrated with an upper bound $\sqrt{T_{\text{cm}}^2-m_{Z'}^2}$. The integral becomes
\begin{align}
	\int_0^{\sqrt{T_{\text{cm}}^2-m_{Z'}^2}} dk \frac{\lvert\vec{k}\rvert^2}{\omega^2 } \left(1-\frac{\lvert\vec{k}\rvert^2}{3\omega^2}\right) &=  \int_0^{\sqrt{T_{\text{cm}}^2-m_{Z'}^2}} dk \frac{\lvert\vec{k}\rvert^2}{k^2 +m_{Z'}^2} \left(1-\frac{\lvert\vec{k}\rvert^2}{3(k^2+m_{Z'}^2)}\right) \\
	&= \frac{T_{\text{cm}}}{2} I(\frac{m_{Z'}}{T_{\text{cm}}}),
\end{align}
with 
\begin{align}
	I(x) \equiv \frac{4}{3}(1-\frac{x^2}{4})\sqrt{1-x^2}-x\tan^{-1}(\frac{\sqrt{1-x^2}}{x}).
\end{align}
Therefore, the emissivity can be finally written in
\begin{align}
	\dot{\epsilon} 	&=  \frac{n_n n_p \alpha\epsilon^2 }{(\pi M T)^{3/2}} \int dT_{\text{cm}} e^{-T_{\text{cm}}/T} T_{\text{cm}}^3 \sigma^{(2)}_{np}(T_{\text{cm}}) I(\frac{m_{Z'}}{T_{\text{cm}}}).
\end{align}
From the equations above, the emissivity is given by
\begin{align}
	\dot{\epsilon} &\equiv \int \int \frac{d^3 k}{(2\pi)^3} \omega d\Gamma_{\text{brem}}^{\text{SRA}} \\
	&=  \frac{2n_n n_p \alpha\epsilon^2 }{(\pi M T)^{3/2}} \int\int dk \frac{\lvert\vec{k}\rvert^2}{\omega^2 } \left(1-\frac{\lvert\vec{k}\rvert^2}{3\omega^2}\right) dT_{\text{cm}} e^{-T_{\text{cm}}/T} T_{\text{cm}}^2 \sigma^{(2)}_{np}(T_{\text{cm}}).
\end{align}
By changing the variable from $k\rightarrow \omega$, the integration over momentum $k$ can be expressed in
\begin{align}
	\int_0^{\sqrt{T_{\text{cm}}^2-m_{Z'}^2}} dk \frac{\lvert\vec{k}\rvert^2}{\omega^2 } \left(1-\frac{\lvert\vec{k}\rvert^2}{3\omega^2}\right) &=  \int_0^{\sqrt{T_{\text{cm}}^2-m_{Z'}^2}} dk \frac{\omega^2-m_d^2}{\omega^2 } \left(1-\frac{\omega^2-m_d^2}{3\omega^2}\right) \\
	&= \int_0^{\sqrt{T_{\text{cm}}^2-m_{Z'}^2}} d\omega \frac{\omega}{\sqrt{\omega^2-m_d^2}} \frac{\omega^2-m_d^2}{\omega^2 } \left(1-\frac{\omega^2-m_d^2}{3\omega^2}\right) \\
	&= \int_{m_d}^{T_{\text{cm}}} d\omega \frac{\sqrt{\omega^2-m_d^2}}{\omega} \left(1-\frac{\omega^2-m_d^2}{3\omega^2}\right) .
\end{align}
Furthermore, the double integral
\begin{align}
	\int_{m_d}^{\infty}\int_{m_d}^{T_{\text{cm}}}d\omega dT_{\text{cm}}... &= \int_{m_d}^{\infty}\int_{\omega}^{\infty} dT_{\text{cm}} d\omega ... 
\end{align}
Therefore, the loop-Bremsstrahlung emissivity spectrum $d\dot{\epsilon}/d\omega$ is
\begin{equation}
	\frac{d\dot{\epsilon}}{d\omega} = \frac{2n_n n_p \alpha\epsilon^2 }{3(\pi M T)^{3/2}}  \sqrt{1-\left(\frac{m_d}{\omega}\right)^2} \left[ 2+\left(\frac{m_d}{\omega}\right)^2 \right] \int_{\omega}^{\infty} dT_{\text{cm}} e^{-T_{\text{cm}}/T} T_{\text{cm}}^2 \sigma^{(2)}_{np}(T_{\text{cm}}).
\end{equation}

\section{Inverse loop bremsstrahlung}
In the soft limit, the only difference between the production and absorption rate in the loop Bremsstrahlung process is the number of initial state polarization combinations. $d\Gamma_{\text{brem}}^{\text{SRA, abs}}$ gain an extra factor $1/3$ during the the polarization average in the amplitude squared comparing to $d\Gamma_{\text{brem}}^{\text{SRA}}$. Recall that the production rate of $Z'$ is given by Eq. \eqref{eq.bremevrate}, it is related to the absorption rate $d\Gamma_{\text{brem}}^{\text{SRA, abs}}$ by
\begin{align}
	d\Gamma_{\text{brem}}^{\text{SRA, abs}} &= \frac{d\Gamma_{\text{brem}}^{\text{SRA}}}{3} \nonumber\\
	&=  \frac{n_n n_p \alpha\epsilon^2 \sqrt{4\pi} }{3\omega \sqrt{M} T^{3/2}} dT_{\text{cm}} e^{-T_{\text{cm}}/T} (\epsilon^\mu J_\mu)^2 T_{\text{cm}}  \frac{d\sigma_{np}}{d\theta_{\text{cm}}} d\theta_{\text{cm}} \nonumber\\
	&=  \frac{4n_n n_p \alpha\epsilon^2 \sqrt{\pi} }{3\omega^3 (M T)^{3/2}} dT_{\text{cm}} e^{-T_{\text{cm}}/T} T_{\text{cm}}^2 \left(1-\frac{\lvert\vec{k}\rvert^2}{3\omega^2}\right) \sigma^{(2)}_{np}(T_{\text{cm}})
\end{align} 
In which $\sigma^{(2)}_{np}(T_{\text{cm}})$ is a shorthand notation defined by
\begin{align}
	\sigma^{(2)}_{np}(T_{\text{cm}}) &\equiv \int_{-1}^{1} (1-\cos\theta_{\text{cm}})\frac{d\sigma_{np}}{d\cos\theta_{\text{cm}}} d\cos\theta_{\text{cm}}
\end{align}

The mean free path $\lambda_{\text{att}}$ is simply related to the absorption rate according to
\begin{align}
	\frac{1}{\lambda_{\text{att}}} &= \frac{\Gamma_{\text{brem}}^{\text{SRA, abs}}}{v} =  \frac{\Gamma_{\text{brem}}^{\text{SRA, abs}}}{\sqrt{1-(m_{Z'}/\omega)^2}} \nonumber\\
	&= \int \frac{4n_n n_p \alpha\epsilon^2 \sqrt{\pi} }{3\omega^3 (M T)^{3/2}} dT_{\text{cm}} e^{-T_{\text{cm}}/T} T_{\text{cm}}^2 \left(\frac{2}{3}+\frac{m_{Z'}^2}{3\omega^2}\right) \sigma^{(2)}_{np}(T_{\text{cm}})/v \nonumber\\
	&= \frac{4n_n n_p \alpha\epsilon^2 \sqrt{\pi} }{3\omega^3 (M T)^{3/2}} \frac{1}{3}\left( \frac{2+(m_{Z'}/\omega)^2}{ \sqrt{1-(m_{Z'}/\omega)^2} } \right)  \int dT_{\text{cm}} e^{-T_{\text{cm}}/T} T_{\text{cm}}^2  \sigma^{(2)}_{np}(T_{\text{cm}})
\end{align}
In which $v$ is the speed of $Z'$. Above formula can be written in a simpler form by defining a dimensionless variable $x\equiv T_{\text{cm}}/T$. The mean free path of the inverse loop Bremsstrahlung process is 
\begin{equation}
	\frac{1}{\lambda_{\text{att}}} = \frac{8n_n n_p \alpha\epsilon^2 }{9\pi\omega^3 } \left( \frac{\pi T}{M}\right)^{3/2} \left( \frac{2+(m_{Z'}/\omega)^2}{ \sqrt{1-(m_{Z'}/\omega)^2} } \right)  \frac{1}{2}\int_0^\infty dx e^{-x} x^2  \sigma^{(2)}_{np}(x)
\end{equation}

\section{The Plateau Phenomenon}
\begin{figure}[htbp]
	\centering
	\begin{overpic}[width=\columnwidth]{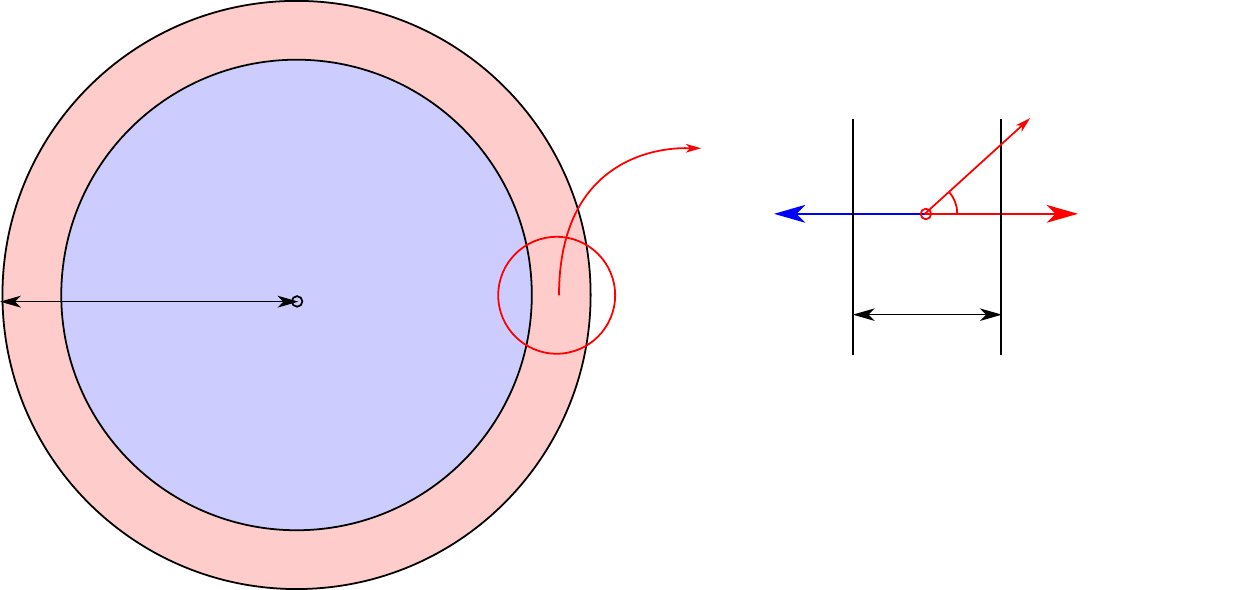}%,grid,tics=10
		\put (1,24) {neutrino sphere $R_\nu$}
		%\put (69,43) {\color{mygreen} $\vec{r}$}
		%\put (30,63) {\color{mylightblue} $\vec{r}\,'$}
		\put (45,41) {\parbox{9em}{$Z'$ within this region can have a chance to escape the sphere.}}
		\put (77,30.5) {\color{red}$\theta$}
		\put (83,37) {\color{mypurple}$1/2$}
		\put (76,42) {\color{mypurple}\parbox{8em}{parallel to the outgoing direction}}
		\put (87,30) {\color{red}$1/2$}
		\put (84.5,26) {\color{red}outgoing}
		\put (72,19) {$\lambda_{\text{att}}$}
		\put (45,9) {\parbox{0.4\linewidth}{
			\begin{align*}
				&\quad L_{\infty}(m_{Z'}) \\
				&=4\pi R^2_\nu\int_{m_{Z'}}^{\infty}\frac{\lambda_{\text{att}}(m_{Z'},R_\nu,\omega)}{{\color{red}2}\times {\color{mypurple}2}}\frac{d\dot{\epsilon}}{d\omega}(m_{Z'},R_\nu,\omega)d\omega \\
				&= \pi R^2_\nu\int_{m_{Z'}}^{\infty}\lambda_{\text{att}}(m_{Z'},R_\nu,\omega)\frac{d\dot{\epsilon}}{d\omega}(m_{Z'},R_\nu,\omega)d\omega
			\end{align*}
		} }
	\end{overpic}
	\caption{A pictorial demonstration for the derivation of $L_{\infty}(m_{Z'})$.}%
	\label{fig.Sphere_e}%
\end{figure}

To understand the consequences of taking $R_{\rm far}=R_\nu$ in the upper integration limit of both emissivity and attenuation calculation, let us consider the case of a huge coupling such that $\lambda_{\rm att}(\omega)\ll R_{\nu}$.
%, i.e., $g_{Z'} \gg 1$. 
Since the interaction length decreases as the coupling constant increases, the new gauge boson $Z'$ produced deep inside the neutrinosphere will have less chance of escaping the sphere than those produced near the surface. %If the coupling constant keeps increasing, eventually 
Generally speaking, only those $Z'$ produced at a thin shell of the thickness of $\lambda_{\rm att}(\omega)$ have the chance to leave the neutrinosphere and contribute to the luminosity. %$Z'$ with energy $\omega$ produced within a shell volume with thickest $\lambda_{\text{att}}(\omega)$ will have a chance to leave the neutrino sphere. 
Since $Z'$ bosons produced within this thin shell can propagate either out or into the neutrinosphere, only a half of the total number of $Z'$ might contribute to the luminosity. Lastly, projecting the trajectory of those particles to ensure that they are propagating outward, we have the luminosity $\Delta L_{\infty}(m_{Z'})$ for $Z'$ in a small energy range $\Delta \omega$
\begin{align}
	\Delta L_{\infty}(m_{Z'},\omega) &= \pi R^2_\nu \lambda_{\text{att}}(m_{Z'},R_\nu,\omega)\frac{d\dot{\epsilon}}{d\omega}(m_{Z'},R_\nu,\omega)\Delta\omega
\end{align}
The overall luminosity $L_{\infty}(m_{Z'})$ can be calculated by sum over all energy contribution. Integrating the above equation with respect to all possible energy $\omega$ yields
\begin{align}
	L_{\infty}(m_{Z'}) &= \pi R^2_\nu\int_{m_{Z'}}^{\infty}\lambda_{\text{att}}(m_{Z'},R_\nu,\omega)\frac{d\dot{\epsilon}}{d\omega}(m_{Z'},R_\nu,\omega)d\omega \label{eq.Linf}
\end{align}
A pictorial demonstration for the derivation of $L_{\infty}(m_{Z'})$ above is given in Fig.~\ref{fig.Sphere_e}. Clearly $L_{\infty}(m_{Z'})$ will tend to a given value according to Eq.~\eqref{eq.Linf} instead of monotonically decreasing down to zero in the large $g_{Z'}$ limit. This is true since $\lambda_{\rm att}(\omega)$ scales as $g_{Z'}^{-2}$ while $d\dot{\epsilon}/d\omega$ scales as $g_{Z'}^{2}$.
It is important to note that $L_{\infty}(m_{Z'})$ given by Eq.~(\ref{eq.Linf}) agrees completely with our full numerical result of $L_{Z'}(m_{Z'})$ in the large $g_{Z'}$ limit. 
%The result will be presented in the next section.

It is interesting to note that the averaging factor $1/4$ appearing in Fig.~\ref{fig.Sphere_e} can be understood as 
$\int_0^1 \cos\theta d(\cos\theta)/\int_{-1}^{1} d(\cos \theta)=1/4$. On the other hand, if only two $Z^{\prime}$ trajectories corresponding to $\theta=0$ and $\theta=\pi$ are considered for evaluating the $Z^{\prime}$ luminosity~\cite{Chang:2016ntp, Croon:2020lrf, Lucente:2020whw}, the averaging factor appearing in $L_{\infty}(m_{Z'})$ would become $1/2$ rather than $1/4$.
%Note that a similar work has been done by McDermott {\it et al.}\cite{Chang}\cite{McDermott} based on a supernova simulation SFHo18.8 provided by Thomas Janka {\itshape et al.} \cite{Bollig:2020xdr}. However, they did not take both the production and absorption/decay processes of $Z'$ on an equal footing. The radius of the spherical volume corresponds to the absorption region $R_{a}$ of the new gauge boson $Z'$ was taken as an arbitrary value that was greater than the one for the production region, i.e., $R_{\text{far}}>R_\nu$ (we use the symbol $R_{a}$ to represent $R_{\text{far}}$ in their work). Hence, the production rate of the new gauge boson $Z'$ was undoubtedly underestimated. To point out this, we also adopted the supernova simulation SFHo18.8\cite{Bollig:2020xdr} here as a comparison. In addition, we treated the $Z'$ production and absorption/decay on an equal footing, i.e., $R_{\text{far}}=R_\nu$, and let the numerical integration to handle the competition between them. 

%As a short summary, we provided the luminosity calculation formula for the new gauge boson $Z'$ produced in the neutrinosphere. It can be calculated by simply inserting the formula of those emissivity spectra (see section \ref{sec.emissivity}) and attenuation length (see section \ref{sec.attenuationlength}) into Eqs.~\eqref{eq.LuminosityZ'} and \eqref{eq.Abar}. 

\end{document}